\documentclass[usenatbib]{mnras}
\usepackage{newtxtext,newtxmath}

\usepackage[T1]{fontenc}

\DeclareRobustCommand{\VAN}[3]{#2}
\let\VANthebibliography\thebibliography
\def\thebibliography{\DeclareRobustCommand{\VAN}[3]{##3}\VANthebibliography}

\usepackage{graphicx}	
\usepackage{amsmath}	
\usepackage{booktabs}
\usepackage{setspace}
\usepackage{ulem}

\newcommand{\paa}{\rm\,Pa$\alpha$}
\newcommand{\lya}{\rm\,Ly$\alpha$}
\newcommand{\ha}{\rm\,H$\alpha$}

\newcommand{\oiii}{\rm\,[O{\sc iii}]}
\newcommand{\oii}{\rm\,[O{\sc ii}]}
\newcommand{\nii}{\rm\,[N{\sc ii}]}

\newcommand{\hii}{\rm\,H{\sc ii}}

\usepackage{color}

\title[Enhanced SF activity in a $z\sim2$ protocluster]{Star formation activity of low-mass galaxies at the peak epoch of galaxy formation probed by deep narrow-band imaging}

\author[K. Daikuhara et al.]{
Kazuki Daikuhara,$^{1}$\thanks{E-mail: daikuhara@astr.tohoku.ac.jp}
Tadayuki Kodama,$^{1}$
Jose M. Pérez-Martínez,$^{1,2,3}$
Rhythm Shimakawa,$^{4,5}$
\newauthor
Tomoko L. Suzuki,$^{6}$
Ken-ichi Tadaki,$^{7}$
Yusei Koyama,$^{8}$
Ichi Tanaka$^{8}$
\\
$^{1}$Astronomical Institute, Tohoku University, 6-3, Aramaki, Aoba, Sendai, Miyagi, 980-8578, Japan.\\
$^{2}$Instituto de Astrofísica de Canarias (IAC), E-38205, La Laguna, Tenerife, Spain.\\
$^{3}$Departamento Astrofísica, Universidad de La Laguna, E-38206 La Laguna, Tenerife, Spain\\
$^{4}$Waseda Institute for Advanced Study (WIAS), Waseda University, 1-21-1, Nishi Waseda, Shinjuku, Tokyo 169-0051, Japan.\\
$^{5}$Center for Data Science, Waseda University, 1-6-1, Nishi-Waseda, Shinjuku, Tokyo 169-0051, Japan.\\
$^{6}$Kavli Institute for the Physics and Mathematics of the Universe (WPI), University of Tokyo, Kashiwa, Chiba, 277-8583, Japan.\\
$^{7}$Faculty of Engineering, Hokkai-Gakuen University, Toyohira-ku, Sapporo 062-8605, Japan\\
$^{8}$Subaru Telescope, National Astronomical Observatory of Japan, 650 North A’ohoku Place, Hilo, HI 96720, USA.
}

\date{Accepted 2024 May 9. Received 2024 May 8; in original form 2023 December 1}

\pubyear{2024}

\begin{document}
\label{firstpage}
\pagerange{\pageref{firstpage}--\pageref{lastpage}}
\maketitle

\begin{abstract}
Low-mass galaxies at high redshifts are the building blocks of more massive galaxies at later times and are thus key populations for understanding galaxy formation and evolution.
We have made deep narrow-band observations for two protoclusters and the general field in COSMOS at $z$ $\sim$ 2. 
In a clumpy young protocluster, USS1558$-$003, at $z$ = 2.53, we find many star-forming galaxies well above the star-forming main sequence of field galaxies at the low-mass end ($M_{\star}/\mathrm{M_{\odot}}<10^{8.9}$). 
This suggests that some environmental effects may be at work in low-mass galaxies in high-density regions to enhance their star formation activities. 
In the core of this protocluster, we also find that enhanced star formation activity of middle-mass galaxies ($10^{8.9} < M_{\star}/\mathrm{M_{\odot}} < 10^{10.2}$) while such trends are not observed in a more mature protocluster, PKS1138$-$262 at $z$ = 2.16.
We expect these activities to be mainly due to galaxy mergers/interactions and differences in the amount of cold gas accretion.
As one piece of evidence, we show that the star formation activity within individual galaxies in the protoclusters is more centrally concentrated than those in the field.
This is probably due to the enhanced interactions between galaxies in the protocluster, which can reduce the angular momentum of the gas, drive the gas towards the galaxy center, and lead to a central starburst. 
\end{abstract}





\section{Introduction}
According to the standard cold dark matter cosmology, the seeds of low-mass galaxies, i.e., small dark matter haloes, are formed by the growth of gravitational instability in the early Universe. 
Those small mass haloes grow gradually to more massive haloes by mergers/accretion.
Therefore, low-mass galaxies at high redshifts are the building blocks of more massive galaxies at later times and are thus key populations for understanding galaxy formation and evolution in general.
However, low-mass galaxies are relatively faint objects and more difficult to observe, and thus their properties are not well understood.

A tight relationship has been established between star formation rate (SFR) and stellar mass over a wide mass range \citep[e.g.][]{Santini2017,Iyer2018}.
These relationships, called star-forming main sequences \citep[hereafter MS;][]{Noeske2007}, have been intensively studied and are well understood \citep[e.g.][]{Brinchmann2004,Daddi2007,Elbaz2007,Noeske2007,Salim2007,Magdis2010,Rodighiero2011,Wuyts2011,Whitaker2012,Whitaker2014,Heinis2014,Schreiber2015,Shivaei2015,Tomczak2016,Santini2017,Iyer2018,Leslie2020,Leja2022,Daddi2022}.
Several large surveys have revealed a number of star-bursting low-mass galaxies ($M_{\star}/\mathrm{M_{\odot}}<10^{9}$) \citep{Maseda2014,Santini2017,Tran2020,Rinaldi2022}.
Most stars in star-bursting low-mass galaxies are thought to be produced by short and intense burst events \citep{Maseda2014,Rinaldi2022}.
Such intense star formation activity in low-mass galaxies is likely to be an important ionization source at high redshifts.

The environment is also an essential parameter in investigating galaxy formation and evolution.
Many studies for nearby galaxies have confirmed that the galaxy properties depend on their environment\citep[e.g.][]{Oemler1974,Dressler1997,Dressler1980,Bamford2009,Peng2010}.
High-density environments such as clusters are expected to develop at the intersection in the cosmic web and undergo accelerated galaxy evolution.
Although the present-day clusters can make an almost negligible contribution to the current cosmic star formation rate density ($\sim$1 percent), protocluster galaxies \citep[the progenitors of present-day cluster galaxies;][for an overview]{Overzier2016} contribute to it significantly  \citep[$\sim$20 percent at $z$ = 2;][]{Chiang2017}.\ Therefore, it is cosmologically important to investigate protoclusters at high redshifts.

The galaxy evolution is expected to proceed in a downsizing fashion from high-mass galaxies to low-mass galaxies as the cosmic age progresses.
Such downsizing effects are seen in various aspects, such as star formation activity \citep{Bell2004,DeLucia2004,DeLucia2006,Kodama2004,Papovich2006,Sparre2015}, morphology \citep{Bundy2006}, chemical evolution \citep{Savaglio2005,vanderWel2005}, and AGN \citep{Hasinger2005}.
It is also known that there is an environmental effect on this downsizing process \citep{Tanaka2005,Tanaka2007,Koyama2007}.
Typical examples of external environmental effects include galaxy mergers/interactions, ram pressure, and strangulation.
The environmental variations that we are witnessing are likely to be produced by a combination of these multiple effects.
Although it is essential to separate these effects and identify their relative importance, it is not well understood when,
where, and how these environmental effects are actually at work at high redshifts to establish environmentally dependent galaxy properties as we see in the present-day Universe.

In order to understand galaxy formation and evolution, it is therefore essential for us to make a high-$z$ galaxy sample covering wide ranges both in mass and environment, and compare galaxy properties in detail such as star formation rate and morphology, as a function of mass and environment.

Based on a deep \ha\ narrow-band (NB) imaging, \cite{Hayashi2016} have discovered some low-mass starbursting galaxies in the mass range of
10$^{8.0-9.3}$ M$_{\odot}$, in one of the spectroscopically confirmed protoclusters, USS1558-003 at $z$ = 2.53 \citep{Hayashi2016}.
However, comparably deep general field survey of low-mass \ha\ emitters at similar redshifts ($z>2$) did not exist at that time,
and therefore we did not know whether the abundance of such low-mass starbursting galaxies were unique to dense protocluster environments or they are more common irrespective of surrounding environments.
We also note that the slope and the absolute value of the main sequence vary depending on the selection method and the tracers used in the survey of star-forming galaxies.
Therefore, it is essential to make a fair comparison sample for the general field based on similar methods and tracers.

With such a motivation, we have performed deep NB imaging observations for the general field, using the same method as used for protoclusters in \cite{Hayashi2016}, to directly make a comparison and reveal, in particular, the environmental effects in low-mass galaxies that are the least understood.
The MS is thought to be produced by a complicated mix of various physics, including mass accretion onto the halos, feedback from star formation and AGN activities, and gas outflows.
Our key question here is whether the MS also depends on environment at the cosmic noon as those processes are likely dependent on surrounding environments.
Previous investigations of the environmental effects at high redshifts have suggested that massive galaxies seem to show a common MS with no environmental dependence \citep{koyama2013,Hayashi2016,Polletta2021,Shi2021,Sattari2021,Jose2023}.
On the other hand, the environmental dependence of the MS is also reported in \citep{Alberts2014,Shimakawa2018uss,Monson2021,Lemaux2022,Jose2024}.
We must extend such investigation to lower-mass galaxies because the physical processes at work in high-mass and low-mass galaxies are different.
Therefore, this work aims to reveal the role of the environment by investigating the star-forming galaxies in the field
especially at low-mass and comparing with those in the protocluster galaxies with the same mass.

This paper is organized as follows. 
Section 2 introduces our project, the original MAHALO survey, and the MAHALO Deep survey.
Section 3 describes our selection of star-forming galaxies and the methods to calculate their physical quantities.
In Section 4, we analyze the environmental effects of galaxies based on luminosity function, stellar mass function, star formation rate, dust extinction, and size.
Section 5 discusses the environmental effects inferred from our results.
Section 6 summarizes the study.
We assume cosmological parameters of $\mathrm{H_0}$ = 70 km s$^{-1}$ Mpc$^{-1}$, $\Omega_{\mathrm{M}}$ = 0.3, and $\Omega_{\Lambda}$ = 0.7.
All the magnitudes presented in this paper are in the AB system \citep{Oke1983}.
Stellar masses and SFRs are estimated assuming the Chabrier initial mass function \citep[IMF;][]{Chabrier2003}.

\section{MAHALO Deep Field/Cluster survey}
Mahalo-Subaru project \citep[][]{Kodama2013} is a systematic NB imaging program of well-known clusters/protoclusters and some general fields using the Multi-Object Infrared Camera and Spectrograph \citep[MOIRCS;][]{Ichikawa2006, suzuki2008} on Subaru Telescope and many customized NB filters installed in the camera (S10B-028I, Kodama et al.; S14B-013, Tadaki et al.).
In particular, we have conducted deeper NB imaging for two protoclusters, namely USS1558-0003 \citep[][]{Hayashi2016,Shimakawa2018uss} (hereafter USS1558) and PKS1138–-262 \citep[SpiderWeb protocluster;][]{Shimakawa2018pks} (hereafter PKS1138) in order to investigate the galaxies in these protoclusters over a wider range in stellar mass and SFR (MAHALO Deep Cluster Survey).
USS1558 is expected to be a kinematically younger protocluster than PKS1138 because USS1558 has a clumpy structure, while PKS1138 has a larger and rounder structure.
Also, the Sunyaev-Zeldovich effect (SZE) signal has been detected for the PKS1138 protocluster by ALMA/ACA observations \citep{Luca2023}.
As already mentioned in the introduction, we have found an intriguing result that the protocluster USS1558 seems to host enhanced SF activities, which could be due to some environmental effects \citep{Hayashi2016,Jose2024}. 
However, in that paper, we could not make a fair comparison with the field galaxies due to the lack of comparably deep NB imaging survey for the field.
Therefore, we have now conducted deep NB imaging for the CANDELS fields in COSMOS and GOODS-S, which we report in this paper, and make a comparison with the protocluster galaxies (MAHALO Deep Field Survey).
In this section, we introduce our MAHALO Deep Field Survey and also briefly describe our previous MAHALO Deep Cluster Survey used for comparisons \citep[][in detail]{Shimakawa2018uss,Shimakawa2018pks}. 

\begin{table*}
    \centering
    \caption{Summary of the MAHALO Deep Feild Survey. The the 3$\sigma$ limiting magnitude is shown estimated with a random 1.5 arcsec diameter aperture photometry of the background field.}\
    \label{tab:filter}
    \begin{tabular}{cccccc|ccc}\hline
    Field & Filter & $\lambda_{\mathrm{center}}$ & $\Delta \lambda_{\mathrm{filter}}$ & $\Delta z$ &  $\delta v$ & Net integration & Liming mag. & FWHM\\
    &&[$\mu$m]&[\AA] && [km] & [hr] & (3$\sigma$, $1^{\prime\prime}.5$)& [arcsec] \\
    \hline
    COSMOS-F1 & NB2095 & 20930 & 260 & 0.04 & 1870 & 8.2 & 24.6 & 0.75\\
    COSMOS-F2 & NB2095 & 20930 & 260 & 0.04 & 1870 & 9.3 & 25.2 & 0.75\\
    \hline
    \end{tabular}
\end{table*}

\subsection{MAHALO Deep Field Survey}
The MAHALO Deep Field Survey is intended to reveal the properties of low-mass and low-SFR galaxies and investigate environmental variations by comparing with those in higher density regions.
We conduct deep NB surveys for the two fields, COSMOS and GOODS-S, with two pointings each to cover ($7^{\prime} \times 7^{\prime}$) field with MOIRCS \citep[][]{Ichikawa2006, suzuki2008} on Subaru Telescope.
These data were obtained on December 11, 2016, and January 3 -- 5, 2018 (S16B-081, S17B-002; PI, Tadayuki Kodama).
The COSMOS field is observed with the NB2095 filter ($\lambda_{\mathrm{center}}$ = 2.093 $\mu$m, $\Delta\lambda$ = 0.026  $\mu$m), while the GOODS-S field is observed with NB2315 filter ($\lambda_{\mathrm{center}}$ = 2.317 $\mu$m, $\Delta\lambda$ = 0.026 $\mu$m).
However, we do not use the GOODS-S data in this work since the GOODS-S data are not deep enough for our analysis due to bad weather during the observations.

The NB filter technique is suitable and powerful for detecting star-forming galaxies in the corresponding redshift slices because the FWHM of the filter is narrow enough to have high sensitivity to emission lines.
The great advantage of the NB survey is that we can not just detect emission line galaxies to a certain line flux limit in an unbiased way but can also obtain emission line flux, which can be converted to SFR, only with imaging observations (NB and BB imaging).
Thus, derived line fluxes are known to have good consistency with the spectroscopic measurements \citep{Shimakawa2015}.
The NB filter, NB2095, corresponds to the emission line redshifts of $z$ = 2.19 (\ha),  $z$ = 3.18 (\oiii), and $z$ = 4.62 (\oii), respectively.
These emission lines (especially \ha) are less affected by dust and allow for a more unbiased sample selection than \lya, which has often been used for low-mass star-forming galaxy surveys.
Moreover, the relatively wide field that MOIRCS can offer with an 8-m class telescope, this survey is efficient in constructing the sample of low-mass star-forming galaxies at the cosmic noon.

The data reduction (mask creation, flat subtraction, sky subtraction, bias variation subtraction, distortion correction, mosaic) is performed using the MOIRCS imaging data
pipeline MCSRED \citep{Tanaka2011}\footnote{https://www.naoj.org/staff/ichi/MCSRED/mcsred.html}.
We first create a tentatively co-added image and use it to create a mask.
Following that, we perform another set of data processing using that masked image.
This process is to ensure masking for faint objects which is essential for securely detecting faint emitters.
MOIRCS is equipped with two HAWAII-2 arrays of 2028 $\times$ 2028 pixels, each of which is analyzed independently and finally mosaiced.
The difference in sensitivity between the two detectors is estimated from the zero-point
of the standard star.

Our imaging data set with broad-band filters consists of B, V (Subaru/Suprime-Cam; S-Cam) \citep{Taniguchi2007,Taniguchi2015}, g, r$^{\prime}$, i, z$^{\prime}$, y (Subaru/Hyper Suprime-Cam; HSC) \citep[][]{Aihara2018,Aihara2022}, F125W, F160W, F606W (Hubble Space Telescope; HST/WFC3), F814W (HST/ACS) \citep[][]{Grogin2011,Koekemoer2011}, J, H, Ks (VISTA) \citep[][]{Mccracken2012} and 3.6 $\mu$m, 4.5 $\mu$m (Spitzer/IRAC) \citep[S-COSMOS;][]{Sanders2007}.
We have matched the pixel scales and PSFs of these BB and NB images with IRAF package; \textit{geomap}, \textit{xyxymatch} and \textit{geotran}, and \textit{gauss}.
The PSF size and the pixel scale of the co-added image are 0.75 arcsec and 0.17 arcsec pixel$^{-1}$,  excluding the Spitzer/IRAC data.

Our NB imaging data reach 24.6 magnitude for COSMOS-F1 (10:00:37.3000, +2:15:12.078) and 25.2 magnitude for COSMOS-F2 (10:00:22.6873, +2:15:17.332) at 3$\sigma$ (Table1).
We distribute random 1.5 arcsec diameter apertures (twice the PSF size) in the images after matching the PSF size and pixel scale and measure the background counts and scatter within the apertures to examine the limiting magnitudes.
We fit a Gaussian function to the negative side of the ADU distribution.
As we explain later, the limiting stellar mass is $\sim10^{8.5}\ \mathrm{M_{\odot}}$ (see \S\ref{Results:3}).
We have reached a low-mass regime that has not been explored in previous NB surveys,
allowing us to make the first fair comparison among different environments at $z\sim2$.

\subsection{MAHALO Deep Cluster survey}
The MAHALO Deep Cluster Survey investigated the over-dense regions around two radio galaxies, USS1558--003 at $z$ = 2.53 and PKS1138--262 at $z$ = 2.16 (S15A-047, Kodama et al.).
Two radio galaxies are massive ($M_{\star}>10^{11-12}\ M_{\odot}$) at $z\sim2$, and expected to become cD galaxies in the future (or already).
Massive galaxies are thought to inhabit high-density regions like the conjunctions of the cosmic web.
In fact, many protoclusters are found using known massive radio galaxies as the landmark  \citep[][for review]{Overzier2016}. 
The following sections describe our data sets used in this paper.
The data reduction has been performed in the same way as the MAHALO Deep Field Survey.
For more information, refer to \cite{Shimakawa2018uss} for USS1558 and \cite{Shimakawa2018pks} for PKS1138.

The data set of the protocluster USS1558 consists of Subaru / S-Cam / B, r$^{\prime}$, and z$^{\prime}$ \citep[][]{Hayashi2012}, HST / WFC3 / F160W \citep[][]{Hayashi2016}, Subaru / MOIRCS / J, H, and  Ks \citep[][]{Hayashi2012,Hayashi2016}, Subaru / MOIRCS / NB2315 \citep[][]{Hayashi2012,Hayashi2016,Shimakawa2018uss}, and Spitzer / IRAC / 3.6 $\mu$m, 4.5 $\mu$m.
The narrow band filter, NB2315 ($\lambda_{\mathrm{center}}$ = 2.093 $\mu$m, $\Delta\lambda$ = 0.26  $\mu$m), can capture the \ha\ emission line at $z$ = 2.53.
The PSF size and pixel scale are $0.7$ arcsec and $0.117$ arcsec pixel$^{-1}$, respectively, excluding the Spitzer/IRAC data. 

The data set of the protocluster PKS1138 consists of Subaru / S-Cam / B and z$^{\prime}$ \citep[][]{koyama2013}, HST / ACS / F814W and F475W \citep[][]{Miley2006}, VLT / HAWK-I / Y, H, and Ks \citep[][]{Miley2006,Dannerbauer2014}, Subaru / MOIRCS / J, and Ks \citep[][]{koyama2013}, Subaru / MOIRCS / NB2071 \citep[][]{koyama2013,Shimakawa2018pks}, and Spitzer / IRAC / 3.6 $\mu$m, 4.5 $\mu$m \citep[][]{Seymour2007}.
NB2071 ($\lambda_{\mathrm{center}}$ = 2.093 $\mu$m, $\Delta\lambda$ = 0.26  $\mu$m) can capture the \ha\ emission line at $z$ = 2.15.
The PSF size and pixel scale are $0.7$ arcsec and $0.117$ arcsec pixel$^{-1}$, respectively, excluding the Spitzer/IRAC data. 
We have two kinds of Ks-band images, one taken with VLT ($\mathrm{Ks\_VLT}$) and the other with Subaru Telescope ($\mathrm{Ks\_Subaru}$).
The $\mathrm{Ks\_VLT}$ image is deeper than $\mathrm{Ks\_Subaru}$ image but the $\mathrm{Ks\_VLT}$ image does not cover the whole region.
For the region where $\mathrm{Ks\_VLT}$ is lacking, we use the $\mathrm{Ks\_Subaru}$ image.
Hereafter, we refer to them commonly as Ks.

In this work, we re-construct the photometric catalogs for both clusters and the field coherently to make a fair comparison between them.
Also, this work differs from the previous one \citep{Hayashi2016,Shimakawa2018uss} because we now consider emission line contributions to the broad-band photometries to achieve more robust spectral energy distribution (SED) fitting, and also exclude low-$z$ contaminant galaxies such as \paa\ emitters based on the color--color diagrams.

\section{Sample selection \& Analysis}

\begin{figure*}
    \includegraphics[width=175mm]{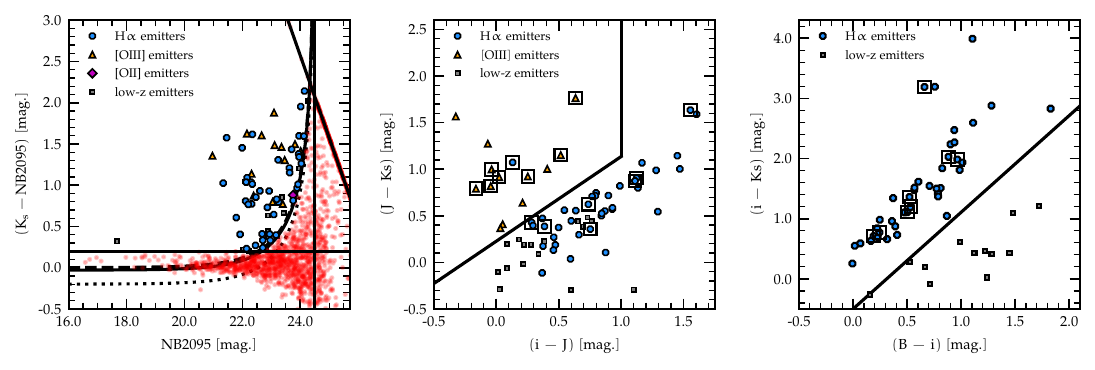}
    \vspace{-0.2cm}
    \caption{The color--magnitude and the color--color diagrams of COSMOS-F1. The blue circles represent the \ha\ emitters, the yellow triangles represent the \oiii\ emitters, the magenta diamonds represent the \oii\ emitters, and the gray squares represent the low-z emitters, respectively. The color--magnitude diagram is used to select the objects which show excess in NB flux with respect to the BB flux. The curves in the figure represent the 3-sigma color excess in K$_{\rm s}$$-$NB2095, corrected for the color terms at their respective redshift. The solid line shows the boundary of the \ha\ emitters, the dashed line shows the \oiii\ emitter, and the dotted line shows the emitters corrected for the color term corresponding to the redshift of the \oii\ emitter. The horizontal solid line indicates the EW-cut. The vertical solid line shows the 3$\sigma$ limiting magnitude of NB2095, and the solid diagonal line indicates the 2$\sigma$ limiting magnitude in K$_{\rm s}$. The emitter candidates are classified into various emission lines at different redshifts based on the color--color diagrams shown in the middle and the right panels. Black open squares show spectroscopic confirmed emitters.}
    \label{fig:selection_cosmosf1}
\end{figure*}

\begin{figure*}
    \includegraphics[width=175mm]{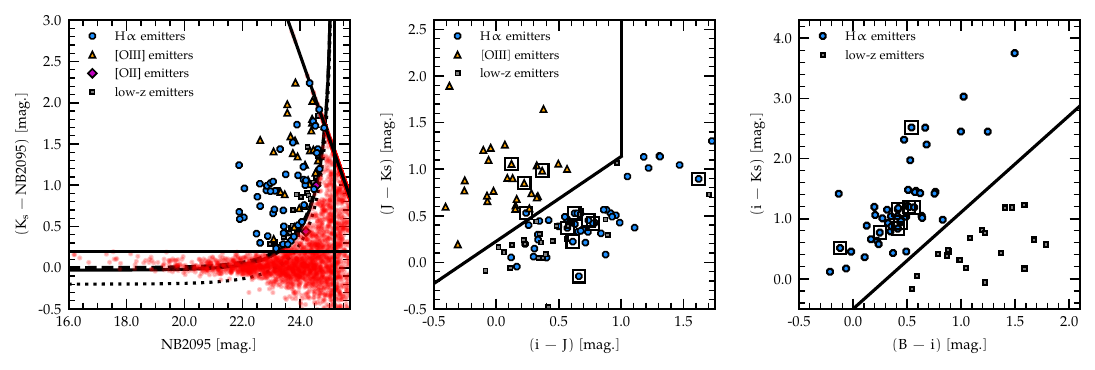}
    \vspace{-0.2cm}
    \caption{The same as Figure~\ref{fig:selection_cosmosf1} but for COSMOS-F2.}
    \label{fig:selection_cosmosf2}
\end{figure*}
 
\subsection{Detection and Photometry}
We perform source extractions from the reduced coadded NB images with SExtractor \citep[version 2.25.2,][]{Bertin1996}.
We also carry out 1.5 arcsec diameter aperture and Kron photometries using the double-imaging mode of the SExtractor.
We basically use the NB image as the detection image.
For IRAC images, however, we conduct independent detections/photometries since the PSF size (FWHM) and pixel scales are very different from the other images.
If the IRAC source position is matched to that of the optical/IR images within one arcsecond, that source is considered to be identical and added to the catalog.
Note that the effect of the PFS size on the blending has not been completely eliminated.
The SExtractor parameters for detections/photometries are set to \textit{DETECT\_MINAREA = 9, DETECT\_THRESH = 1.2, ANALYSIS\_THRESH = 1.2, DEBLEN\_MINCONT = 0.001, BACK\_SIZE= 64, BACK\_FILTERSIZE = 5, BACKPHOTO\_TYPE = LOCAL, BACKPHOTO\_THICK = 32}, and $k=2.5$.

Background noise $\sigma (S)$ was estimated by random photometry with Photoutils \citep{Photutils} following \cite{Bunker1995,Shimakawa2018pks}.
Photutils is a package based on Python code for source extractions and photometry.
We confirmed that the aperture photometry results with Photoutils are consistent with SExtractor output.
Some studies have estimated errors assuming that background noise is proportional to the square root of the area.
However, the assumption leads to an underestimation, as pixels correlate with each other due to pixel scale and PSF matching.
We, therefore, assume that background noise follows $\sigma(S)=C S^{N}$ where $C$ is the proportionality coefficient, $S$ is an aperture area, and $N$ is a slope, and derive these variables for each image.
We perform 1.0 to 3.0 arcsec random diameter aperture photometry in 0.2 increments of 2000 each, then derived $C$ and $N$ values, i.e., the error as a function of the aperture area.

\subsection{Color--magnitude diagram}
\label{Colormag}
We select \ha\ emitters at $z\sim2$ by combining NB and BB imaging and making color--magnitude diagram.
The flux density (magnitude) of an object in NB becomes brighter (smaller) than that in BB if a certain emission line enters into the NB and the BB samples the continuum light underneath the emission line.
The NB filters that we use in this paper (NB2071, NB2095, and NB2315) are all in the Ks band, and thus we can use Ks $-$ NB color versus NB magnitude diagram to select \ha\ emitter candidates.
The following criteria are used for the selection for COSMOS and PKS1138:
\begin{equation}
  m_{\mathrm{Ks}}-m_{\mathrm{NB}}>-2.5\log_{10}\left[1-\frac{\,3\sqrt{\sigma_{\mathrm{Ks}}^2-\sigma^2_{\mathrm{NB}}}}{f_{\mathrm{NB, apr}}}\right] + \xi,
\label{eq:SNcut}
\end{equation}
where $m_{i}$ is an aperture magnitude, $f_{\mathrm{NB, apr}}$ is a NB flux density, and $\xi$ is a color term.
The color term is to correct for a difference in the effective wavelength of the two filters.
We use the \cite{BruzualCharlot2003} simple stellar population (SSP) synthesis models to derive the intrinsic color of star-forming galaxies as a function of redshift with a step of 0.01.
Here, we assume a constant star formation with an age of 0.1 Gyr or 1Gyr, or a delayed star-formation history (SFH) with $\tau$ = 10 Gyr.
The results show that the continuum variation does not affect our conclusion ($|\xi| < 0.03$) for both \ha\ emitters at $z\sim2$ and \oiii\ emitters at $z\sim3$.
For the \oii\ emitters, however, the color term can be significant ($\xi > 0.2$).
In this work, we apply a color term of $\xi$ = $-0.03$ (\ha), 0.001 (\oiii), and $-0.2$ (\oii) for COSMOS, and 0.003 (\ha) and 0.03 (\oiii) for USS1558, and $-0.02$ (\ha) for PKS1138.

After considering these color effects, we select objects
with a color excess of 3$\sigma$ or more as emitter candidates (Equation~\ref{eq:SNcut}).
On top of that,we also apply a constant color cut of Equation~(\ref{eq:EWcut}) and (\ref{eq:EWcutUSS}) in order to securely select emitter candidates above the intrinsic variation of the colors.
For COSMOS and PKS1138, we apply
\begin{equation}
m_{\mathrm{Ks}}-m_{\mathrm{NB, apr}}>0.20,
\label{eq:EWcut}
\end{equation}
while for USS1558, we apply
\begin{equation}
\centering
m_{\mathrm{Ks}}-m_{\mathrm{NB, apr}}>0.275.
\label{eq:EWcutUSS}
\end{equation}
These color thresholds correspond to the limit in the equivalent width (EW) of emission lines, and Equation~(\ref{eq:EWcut}) and (\ref{eq:EWcutUSS}) actually correspond to EW = 30 \text{\AA}.
If the object is detected only below 2$\sigma$ in Ks band, the 2$\sigma$ value is taken as the Ks magnitude.
We note, however, that when we discuss stellar masses, we limit our sample to only those galaxies which are detected in one of the nearby BBs among J, H, and F160W at or above 3$\sigma$.

\subsection{Color--color diagram}
Once we have line emitter candidates, the next step is to separate those emitters into different lines at different redshifts.
For this purpose, we use either spectroscopic redshifts, if available, or color--color diagrams.
The spectroscopic data of VVDS \citep{LeFevre2013}, 3D-HST\citep{Skelton2014,Momcheva2016}, and MOSDEF \citep{Kriek2015} are used for COSMOS.
For USS1558 and PKS1138, we utilize the spectroscopic catalogs from the follow-up observations \citep{Shimakawa2014,Shimakawa2015,Jose2023}.
For the rest of the objects, color--color diagrams are be used to separate different emitters at different redshifts if the SED covers the spectral break features, such as Balmer/4000\AA\ breaks. 
For example, the NB2095 emitters are mainly populated by \ha\ emitters at $z=2.19$, \oiii\ emitters at $z=3.18$, \oii\ emitters at $z=4.62$, and \paa\ emitters at $z=0.116$.
 
For the NB2095 data, \ha\ emitters are selected with the criterion of equation ~(\ref{eq:color_1}), while \oiii\ or \oii\ emitters are selected with the criterion of Equation~(\ref{eq:color_2}).
These criteria are similar to the ones used in \cite{Tadaki2013}, and are confirmed to be effective by spectroscopic observations. 
\begin{equation}
\centering(J-Ks) \geq 0.91 (i-J) + 0.227,\ (i-J) \leq 1.0, \label{eq:color_1}
\end{equation}
\begin{equation}
  (J-Ks) \geq 1.61 (b-i^{\prime}) + 0.227. \label{eq:color_2}
\end{equation}
The \oii\ emitter is defined by the following criteria: 
\begin{equation}
   r - i > 1.2,\quad i -z^{\prime}<0.7,\quad r - i >1.5 (i-z^{\prime})+1.0.
\end{equation}

As a result, we identify in total 78 emitters in COSMOS-F1 (42 \ha\ emitters, 13 \oiii\ emitters, and 1 \oii\ emitter), and 103 emitters in COSMOS-F2 (48 \ha\ emitters, 32 \oiii\ emitters, and 2 \oii\ emitters).
Our final sample consists of 90 \ha\ emitters at $z$ = 2.19, 45 \oii\ emitters at $z$ = 3.18, and 3 \oii\ emitters at $z$ = 4.62.
This paper focuses mainly on \ha\ emitters because we want to compare these field \ha\ emitters directly with those in two protoclusters at $z$ $\sim$ 2 identified by our previous studies \citep{Shimakawa2018uss,Shimakawa2018pks}.
As for \oiii\ emitters at $z$ $\sim$ 3, there are no such protocluster samples for comparison at the moment, and we consider it as a future extension work.

We do not repeat the data processing description of the two protoclusters in this paper, and the readers should refer to \cite{Shimakawa2018uss,Shimakawa2018pks} for details.
The 3$\sigma$ limiting magnitudes (within 1.5 arcsec diameter aperture) of the NB and Ks images of USS1558 reach 24.14 mag and 24.75 mag, respectively.
The \ha\ emitters in USS1558 are selected so as to satisfy the following two criteria;
\begin{equation}
(r^{\prime} - J) < 0.9,  \qquad (J - Ks) > (r^{\prime} - J) - 0.6,\label{eq:color_3}
\end{equation}
\begin{equation}
(B - r^{\prime}) < 1.0, \qquad (r^{\prime} - Ks) > 2.5\times(B - r^{\prime})-1.6.\label{eq:color_4}
\end{equation}
The 3$\sigma$ limiting magnitudes (within 1.5 arcsec diameter aperture) of the NB and Ks images of 
PKS1138 reach 23.92 mag and 24.75 mag, respectively.
The \ha\ emitters in PKS1138 are selected so as to satisfy the following two criteria;
\begin{equation}
    (z - Ks) > (B - z) - 0.2,\qquad (z - Ks) > 2.5\label{eq:color_5},
\end{equation}
\begin{equation}
    (J - Ks) > 1.45\,(z - J) - 0.3,\qquad (z - J) > 0.8\label{eq:color_6}.
\end{equation}
As a result, 113 \ha\ emitters (and 15 \oiii\ emitters) and 82 \ha\ emitters are identified in USS1558 and PKS1138, respectively.
We note that the available photometric bands are different among the three environments (USS1558, PKS1138 protoclusters and COSMOS field) although they are similar. Therefore, we are not able to apply exactly the same color criteria.

\begin{figure*}
    \includegraphics[width=177mm]{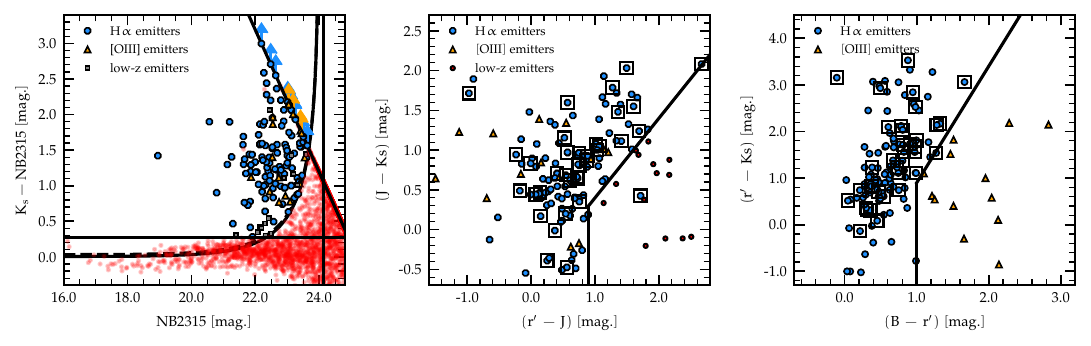}
    \vspace{-0.2cm}
    \caption{Color--magnitude and color--color diagrams of USS1558. The symbols are the same as in Figure~\ref{fig:selection_cosmosf1}.}
    \label{fig:selection_uss}
\end{figure*}

\begin{figure*}
    \includegraphics[width=175mm]{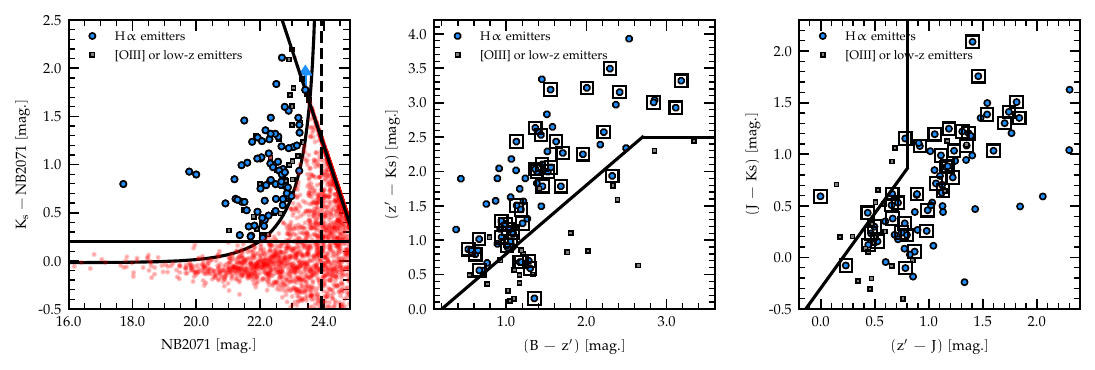}
    \vspace{-0.2cm}
    \caption{Color--magnitude and color--color diagrams of PKS1138. The symbols are almost the same as in Figure~\ref{fig:selection_cosmosf1}. Due to the limitations of our data set, we are unable to classify \oiii\ or low-z emitters. Therefore, in this figure, we show the \oiii\ and low-z emitters as gray squares.}
    \label{fig:selection_pks}
\end{figure*}

\subsection{Completeness}
\label{Completeness}
Completeness is investigated as a function of NB magnitude by Monte Carlo simulations.
Firstly, mock objects are generated with a PSF size of $\sim0.75$ for COSMOS and $\sim0.8$ for USS1558 and PKS1138).
For the Ks image, mock objects are embedded in the reduced image with a Ks magnitude from 20.0 to 26.0 with a step of 0.1 magnitude. 
For the NB image, mock objects are generated in the magnitude range of 20.0 and 26.0 magnitude with a step of 0.1.
We incorporate \ha\ line flux to the NB magnitude to apply color selection (Equation~\ref{eq:SNcut}) for mock objects.
Since there is a correlation between NB magnitude and line flux, the relation of the line flux as a function of NB magnitude is determined and the \ha\ emission line flux is added to NB magnitudes.
The generated mock objects are randomly embedded in the reduced NB and Ks image.
We perform object detections and photometries using the same method for the images with the mock objects.
The recovery rate is then examined by applying the same criteria (Equation~\ref{eq:SNcut}-\ref{eq:EWcutUSS}) to these images.
This procedure is repeated 100 times.
The average recovery rate is described by the following equation;
\begin{equation}
   C(m_{\mathrm{NB}}) = a\times \mathrm{Erfc}\left[\frac{x-b}{c}\right]+d 
   \label{eq:completeness}
\end{equation}
where the variable $a,b,c,d$ are determined.
We calculate the 68\% completeness magnitudes of 23.47 and 23.88 using the fitted Equation~(\ref{eq:completeness}) for COSMOS-F1 and F2, respectively.
For USS1558 and PKS1138, the 68\% completeness magnitudes are 23.06 and 22.45.

We note that the \ha\ emitter selection technique presented with equations~(\ref{eq:SNcut}) -- (\ref{eq:color_6}) is highly dependent on photometric errors. 
In particular, low-mass galaxies tend to have larger photometric errors and thus lower recovery rates.
The difference in the recovery rate affects the results when we calculate the \ha\ luminosity functions and the stellar mass functions.
In this study, we obtain the \ha\ luminosity function and the stellar mass function by applying the completeness corrections using the Equation~(\ref{eq:completeness}).

\subsection{Stellar-mass \& Dust extinction}
We conduct the SED-fitting to derive stellar masses ($M_{\star}$) and dust extinctions with the CIGALE code \citep[Code Investigating GALaxy Emission, ver. 2022.0,][]{Burgarella2005,Noll2009,Boquien2019}.
We use a six-module, \textit{dustatt modified starburstmodule} \citep[][]{Calzetti2000s}, \textit{bc3} \citep{BruzualCharlot2003}, \textit{nebula}, \textit{dale2014}, \textit{Dust Emission}, \textit{redshifting}, produced by CIGALE.

We assume delayed exponentially declining star formation history (SFH) presented by
\begin{equation}
    \mathrm{SFR}\propto t \exp\left(-\frac{t}{\tau}\right),
\end{equation}
where $\tau$ is the timescale of star formation, and it also corresponds to the epoch when the SFR is peaked.
SFR has smoothly decreased afterward toward the present.
For \ha\ emitters, $\tau$ and $t$ are chosen from $10^{9}$ to $10^{11}$ yrs and from $10^{7.6}$ to $10^{9.4}$ yrs, respectively.
The stellar extinction ($A_V$) is chosen from 0 to 2.4 mag, and we assume $E(B-V)_{\mathrm{gas}}=E(B-V)_{\mathrm{stellar}}$.
This assumption inevitably introduces some uncertainties.
\cite{Koyama2019} finds that the ratio $E(B-V)_{\mathrm{gas}}/E(B-V)_{\mathrm{stellar}}$ increases with increasing stellar mass, and decreases with increasing specific SFR.
However, since we consistently use the same assumptions for all galaxies in this study, we can still compare the derived physical quantities such as SFR in a relative sense.
Redshifts of the emitters without spectroscopic redshifts are assumed to be $z$ = 2.19 (COSMOS), $z$ = 2.16 (PKS1138), and $z$ = 2.56 (USS1558), respectively.
Metallicity of the stellar population is fixed to $Z$=0.04, but adding the $Z$=0.008 and 0.02 templates has little impact on our final results.
For more detailed parameters, see Table~\ref{table:cigale}.
Under these settings, CIGALE calculates the reduced-$\chi^{2}$ for each SED model and determines the best-fit model SED with the lowest $\chi^{2}$ value. 
Note that the stellar mass may be overestimated if AGN is present since the light contribution from AGN is not considered.

\begin{table*}
 \caption{Input Parameters for SED Fitting with CIGALE}
 \label{table:cigale}
 \centering
  \begin{tabular}{lcc}
  \toprule
  $\bullet$\quad \textbf{Stellar Population} \cite{BruzualCharlot2003}&&\\
  \midrule
  Initial mass function &  & \cite{Chabrier2003} \\
  Metallicity &  & 0.004 \\
  \midrule
  $\bullet$\quad \textbf{Delayed Star Formation History}&&\\
  \midrule
  e-folding time of the stellar population model & $\tau_{\mathrm{main}}$ & $\log_{10} (\tau_{\mathrm{main}})$ from 9.0 to 11.0 in steps of 0.1. \\
  Age of the stellar population in the galaxy & $\mathrm{age_{main}}$ & $\log_{10} (\mathrm{age_{main}})$ from 7.6 to 9.4 in steps of 0.1. \\
  Mass fraction of the late burst population & $f_{\mathrm{burst}}$ & 0.0 \\ 
  \midrule
  $\bullet$\quad \textbf{Nebular Emission}&&\\
  \midrule
  Ionisation parameter & $\log U$ & -2.4, -2.5, -2.6, -2.7, -2.8\\
  Gas metallicity & zgas & 0.004 \\
  Electron density & $n_e$ & 100\\
  Fraction of Lyman continuum photons escaping the galaxy
  & $f_\mathrm{esc}$ & 0.0 \\
  Fraction of Lyman continuum photons absorbed by dust& $f_\mathrm{dust}$ &  0.0 \\
  \midrule
  $\bullet$\quad \textbf{Dust extinction} \cite{Calzetti2000} &&\\
  \midrule
  The color excess of the nebular lines light & $E(B - V)_\mathrm{lines}$ & $A_V$ from 0 to 3 mag. in steps of 0.1.\footnote{$A_v=E(B - V)_\mathrm{lines}/E(B - V)_\mathrm{stellar}$} \\
  Nebular to continuum ratio & $E(B - V)_\mathrm{factor}$ & 1.0 \\
  & $R_v$ & 4.05\\
  \midrule
  $\bullet$\quad \textbf{Dust Emission} \cite{Dale2014} &&\\
  \midrule
  AGN fraction & $f_{\mathrm{AGN}}$ & 0.0\\
  Power-law slope & $\alpha$ & 2.0\\
  \bottomrule 
  \end{tabular}
\end{table*}

\subsection{Star-formation rate}
One of the great advantages of NB imaging is that we can not only select emission line galaxies but also obtain line fluxes \citep{Bunker1995}.
The observed NB and BB fluxes are respectively expressed as follows;
\begin{equation}
    F_{\mathrm{NB}}\ 
    \simeq\ f_{\mathrm{NB}}\, \Delta_{\mathrm{NB}} \ =\  F_{\mathrm{line}}+f_{\mathrm{con}}\,\Delta_{\mathrm{NB}}\label{eq:nb_flux}
\end{equation}
\begin{equation}
    F_{\mathrm{BB}}\ 
    \simeq\ f_{\mathrm{BB}}\, \Delta_{\mathrm{BB}}\ =\  F_{\mathrm{line}}+f_{\mathrm{con}}\,\Delta_{\mathrm{BB}}\label{eq:bb_flux}
\end{equation}
Solving for $F_{\mathrm{line}}$ using Equation~(\ref{eq:nb_flux}) and (\ref{eq:bb_flux}), we get
\begin{equation}
    F_{\mathrm{line}} = \frac{f_{\mathrm{NB}}-f_{\mathrm{BB}}}{1-\Delta_{\mathrm{NB}}/\Delta_{\mathrm{BB}}}\Delta_{\mathrm{NB}}\label{eq_line}
\end{equation}
\begin{equation}
    f_{\mathrm{con}} = \frac{f_{\mathrm{BB}}-f_{\mathrm{NB}}(\Delta_{\mathrm{NB}}/\Delta_{\mathrm{BB}})}{1-\Delta_{\mathrm{NB}}/\Delta_{\mathrm{BB}}}\label{eq_con}
\end{equation}
where $f_{\mathrm{NB}}$ and $f_{\mathrm{BB}}$ are NB and BB flux densities, respectively, and $\Delta_{\mathrm{NB}}$ and $\Delta_{\mathrm{BB}}$ are FWHMs of NB and BB, respectively. $F_{\mathrm{line}}$ is a line flux, and $f_{\mathrm{con}}$ is a flux density of continuum.
As mentioned in section~\ref{Colormag}, $f_{\mathrm{NB}}$ and $f_{\mathrm{BB}}$ are not sampling exactly the same continuum levels since the effective wavelengths of NB and BB filters are different.
In this study, a colour-corrected Ks flux density $f_{Ks}^{\prime}$ is used to determine the $F_{\mathrm{line}}$ as follows,
\begin{equation}
    F_{\mathrm{line}} = \frac{f_{\mathrm{NB}}-f_{\mathrm{Ks}}^{\prime}}{1-\Delta_{\mathrm{NB}}/\Delta_{\mathrm{Ks}}}\Delta_{\mathrm{NB}}\label{eq:lineflux}.
\end{equation}
We note that as for the \ha\ emission line at $z$ = 2.53 traced by NB2315, the line flux has almost no contribution to the Ks-band flux since it is located at the edge of the Ks-band filter.
Therefore, $F_{\mathrm{line}}$ is derived by the following equation;
\begin{equation}
    F_{\mathrm{line}} = (f_{\mathrm{NB}}-f_{\mathrm{BB}}) \Delta_{\mathrm{NB}}\label{eq_lineuss}
\end{equation}

In order to derive star formation rate from \ha\ emission line strength, 
it is necessary to correct for dust extinctions, both the one directly associated to the target galaxy and the other seen through our Galaxy (Galactic dust extinction).
The former dust extinction for \ha\ emission line emitted from the \hii\ regions can be derived from $E(B-V)_{\mathrm{gas}}$ value obtained from the SED fitting (assuming the nebular to continuum ratio of 1 as in Table 2), and the extinction curve of $A^{\prime}(\lambda)= k^{\prime}(\lambda) E(B-V)_{\mathrm{nebula}}$ \cite{Calzetti2000}.
The coefficient $k$ is defined as follows;\\
in the case of 0.63 $\mu$ m $\leq$ $\lambda$ $\leq$ 2.20 $\mu$ m, 
\begin{equation}
k^{\prime}(\lambda) = 2.659(-1.857+1.040 / \lambda)+R_{V}^{\prime}.
\end{equation}
The Galactic dust extinction is determined using the Extra-galactic Database extinction low calculator\footnote{http://irsa.ipac.caltech.edu/applications/DUST/}.
These dust extinction values are used to derive dust-corrected emission line fluxes;
\begin{equation}
    f_{\mathrm{line,\, corr}} =  f_{\mathrm{line}}\times 10^{0.4A^{\prime}(\lambda)}.
\end{equation}
Furthermore, for \ha\ emitters, it is necessary to remove the \nii\ emission line contribution that simultaneously enters our NB filter.
To estimate the \nii\ line flux, we use the stellar mass--metallicity relation and the N2 index ($N2=\log_{10}$(\nii $\lambda6585$/ \ha)) calibration \citep[][Figure 19]{steidel2014};
\begin{equation}
    N2 = -0.86 + 0.35 (\log_{10}(M_{\star}/\mathrm{M_{\odot}})-10).
\end{equation}
The relationship between the N2 index and the metallicity is adopted from \cite{Pettini2004}.
We then convert the \nii-corrected \ha\ line flux to the intrinsic \ha\ luminosity by the following equation;
$L_{\mathrm{H}\alpha}=4\pi D_L^2\ f_{\mathrm{line,\, corr}}$, where $D_L$ is the luminosity distance.
The SFR ([$\mathrm{M_{\odot}\,yr^{-1}}$]) is then estimated following the \cite{Kennicutt2009};
\begin{equation}
    \mathrm{SFR_{H\alpha}}=4.82\times10^{-42}\left(\frac{f}{1.64}\right)^{-1}\left(\frac{L_{\mathrm{H\alpha}}}{\mathrm{erg\,s^{-1}}}\right),\label{eq_SFRha}
\end{equation}
where $f$ is a conversion factor from the Salpeter IMF \citep{Salpeter1955} to the Chabrier IMF\citep{Chabrier2003}, which we assume to be $f=1.64$ \citep{Madau2014} in this study.

\section{Results}
\subsection{Spatial distribution of \ha\ emitters}
\label{Results:1}
\begin{figure}
	\includegraphics[width=\columnwidth]{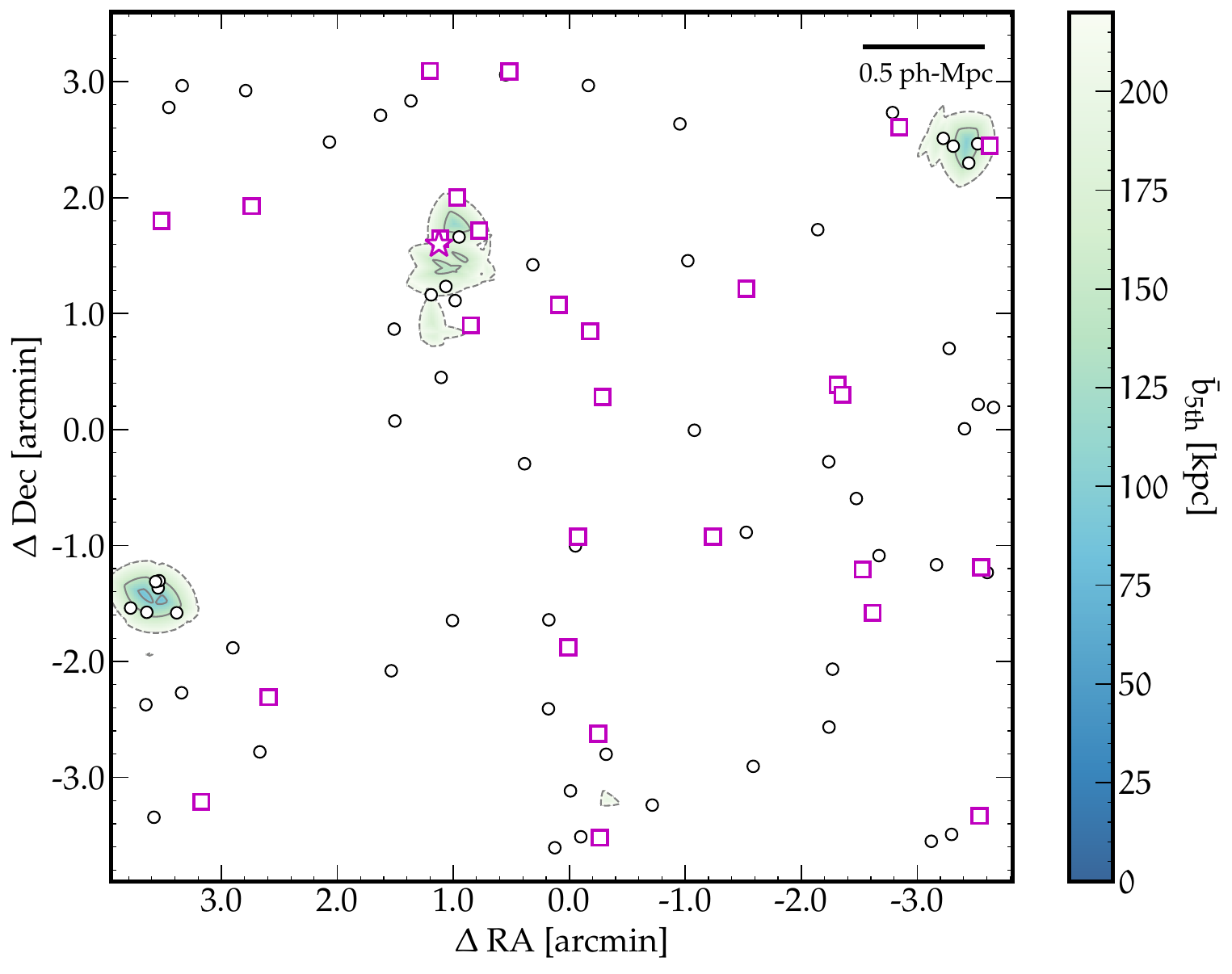}
    \vspace{-0.2cm}
    \caption{Spatial distribution of \ha\ emitters in the COSMOS field. Black circles show the \ha\ emitters with $M_{\star}>10^{9.0}\,M_{\odot}$, while magenta squares and star marks show lower mass galaxies. The magenta squares represent normal star-forming galaxies, while the magenta stars represent star-bursting \ha\ emitters ($\Delta$MS $>$ 0.3).}
    \label{fig:dist_cosmos}
\end{figure}

\begin{figure*}
	\includegraphics[width=12cm]{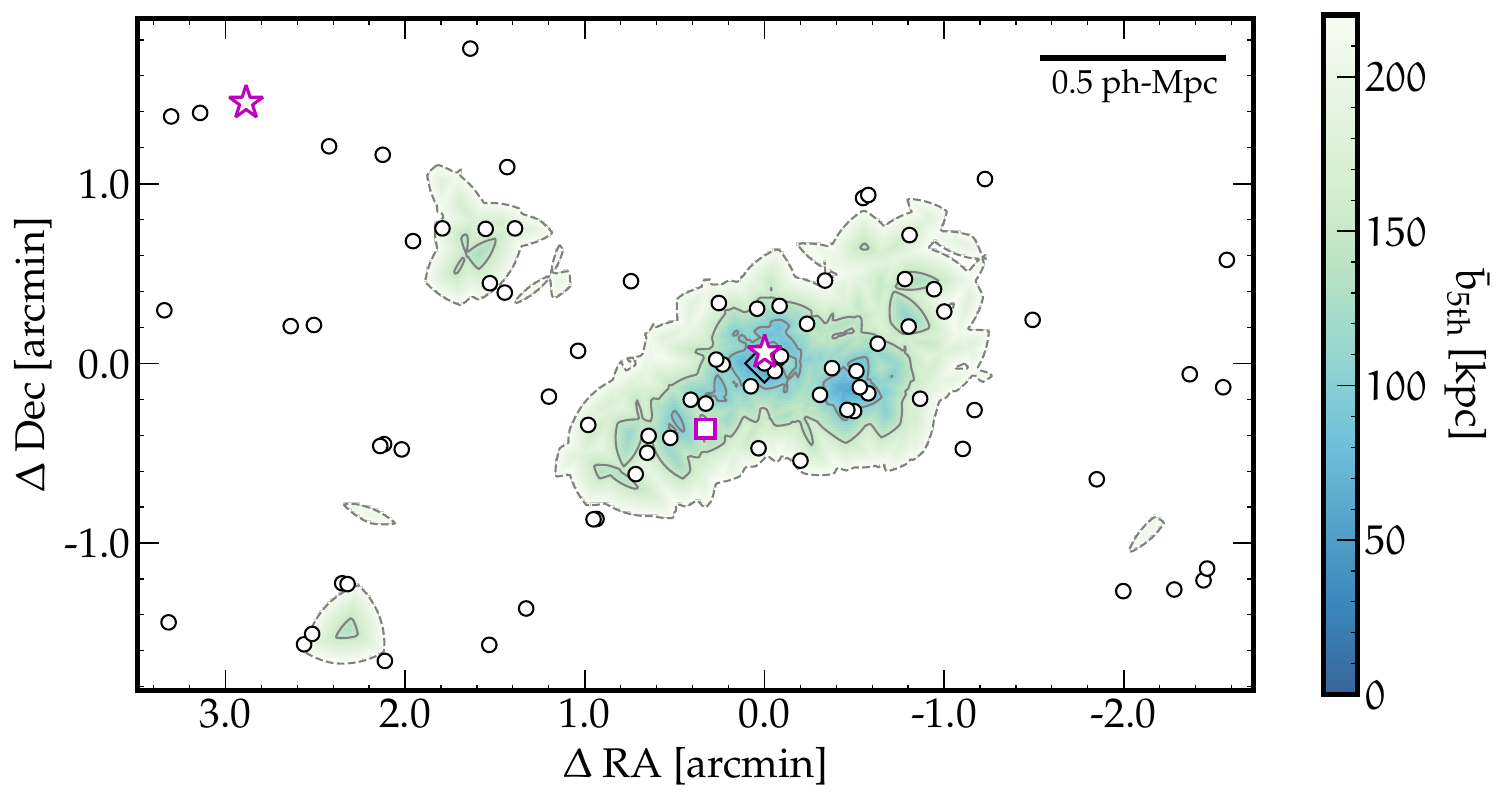}
    \vspace{-0.2cm}
    \caption{Spatial distribution of \ha\ emitters in the PKS1138 protocluster. Meaning of the symbols is the same as  Figure~\ref{fig:dist_cosmos}. Note that the photometric data of PKS1138 is shallower than those of the rest of our target fields; hence few low mass galaxies have been detected. The color scale indicate the mean projected distance, and the contours corresponds to 100, 150, and 220 kpc.  In this paper, the separation between the protocluster core and the outer regions is defined at 220 kpc.}
    \label{fig:dist_pks}
\end{figure*}

\begin{figure}
	\includegraphics[width=\columnwidth]{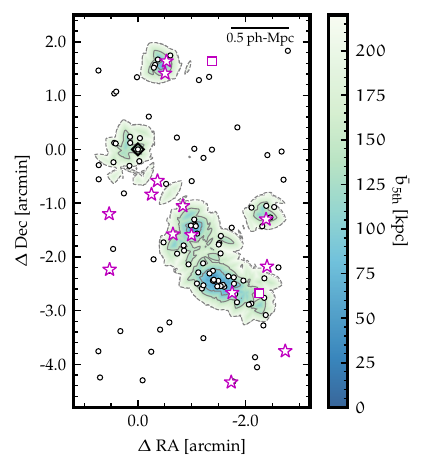}
    \vspace{-0.2cm}
    \caption{The same as Figure~\ref{fig:dist_pks} but for the USS1558 protocluster. USS1558 shows clumpy structures compared to PKS1138 which indicates a more symmetric structure around the central radio galaxy.}
    \label{fig:dist_uss}
\end{figure}

Figure~\ref{fig:dist_cosmos} -- \ref{fig:dist_uss} show the spatial distributions of star-forming galaxies in the field at $z$ = 2.19, PKS1138 protocluster at $z$ = 2.16, and USS1558 protocluster at $z$ = 2.53.
We calculate the mean projected distance $\bar{b}_{\mathrm{5th}}$ to quantify the local environment based on 2-D number density of galaxies  \citep{Shimakawa2018uss}.
$\bar{b}_{\mathrm{5th}}$ show the distance to the fifth nearest \ha\ emiiter.
Note that we are seeing the structures traced only by emission line galaxies, and these may not represent the overall structures.
However, the red galaxies \citep[J - Ks $>$ 1.38 (AB);][]{Franx2003} associated to the two protoclusters are also known to be distributed preferentially along these structures \citep{Hayashi2012,koyama2013,Shimakawa2018uss}.

The spatial distribution of the emitters in the COSMOS field also shows several moderately dense regions.
It is in a good agreement, with only a few exceptions, with the structure traced by $\bar{b}_{\mathrm{5th}}$ based on the photometric redshifts obtained from the COSMOS2020 catalog \citep{Weaver2022}.
We also note that a part of our survey field is located near the known cluster at $z$ = 2.09, which is confirmed by KECK/MOSFIRE spectroscopy \citep[][]{Yuan2014}.
However, it should not be a problem as NB2095 filter can trace only the \ha\ emitters in the narrow redshift slice of $z = 2.19 \pm 0.02$.
In any case, our COSMOS field clearly represents a relatively low-density environment compared to the other two protoclusters regions, as shown in Figure~\ref{fig:dist_cosmos}.
For these reasons, the COSMOS data is used as the field comparison sample in this study.

\subsection{H$\alpha$ luminosity function and stellar mass function}
\label{Results:2}
The \ha\ luminosity ($L_{\mathrm{H\alpha}}$) and stellar mass ($M_{\star}$) function represent the number of galaxies per unit volume per each bin of $L_{\mathrm{H\alpha}}$ and $M_{\star}$.
The number of galaxies in each bin is calculated as follows;
\begin{equation}
    \phi(\log L)=\frac{1}{V}\,\Sigma_{i}\frac{1}{C_i(m_{\mathrm{NB}}) \cdot \Delta(\log L)}
\end{equation}
\begin{equation}
    \phi(\log M)=\frac{1}{V}\,\Sigma_{i}\frac{1}{C_i(m_{\mathrm{NB}}) \cdot \Delta(\log M)},
\end{equation}
where $V$ is the survey volume and $C(m_{\mathrm{NB}})$ is the completeness as a function of NB magnitude.
The completeness correction employs a similar method as in \cite{Shimakawa2018uss,Shimakawa2018pks}, but our method differs in that it also considers the condition of the Ks image.
The survey volumes of COSMOS, PKS1138, and USS1558 are 7179, 3226, and 3118 comoving-Mpc$^3$, respectively.
Note, however, that the actual redshift distributions of the protoclusters are narrower than the redshift range ($\Delta z \sim 0.04$) corresponding to the NB width. 
Thus, the actual volumes of the protoclusters are smaller than the values above.
The errors are estimated as 1$\sigma$ Poisson errors \citep{Gehrels1986} corresponding to 1$\sigma$ of Gaussian statistics.
The fixed bin size is 0.24 for the H$\alpha$ luminosity function and 0.44 for the stellar mass function.

We distinguish and remove some confirmed AGNs from our sample.
We look for matches within 0.5 arcsec from the objects in the COSMOS 3GHz AGN Catalog.
One object are found in the \ha\ emitter (and another object in the \oiii\ emitter).
The depths of these observations have reached $2.2 \times 10^{-16}\ \mathrm{erg\, s^{-1}\, cm^{-2}}$ at 0.5-2 keV, $1.5 \times 10^{-15} \mathrm{erg\, s^{-1} cm^{-2}}$ at 2-10 keV, and $8.9 \times 10^{-16}\ \mathrm{erg s^{-1} cm^{-2}}$ at 0.5-10 keV\citep{civano2016}
In the protocluster USS1558 and PKS1138, we used the AGN sources selected in \cite{macuga2019} and \cite{Shimakawa2018pks}, respectively.
It should also be noted that our AGN samples do not contain an AGN in the \cite{macuga2019} catalog as it is located at the edge of our image and its S/N ratio is lower.
The relationship between high-redshift AGN sources and their environment is still poorly understood.
Our AGN classification is not complete, but it effectively removes some bright AGNs.
We remove these objects when comparing the $L_{\mathrm{H\alpha}}$ function.

The obtained values are then fitted with the \cite{Schechter1976} function using the Markov Chain Monte Carlo (MCMC) method;
\begin{equation}
    \phi(L) \mathrm{d} L=\mathrm{\phi^{*}}\left(\frac{L}{\mathrm{L^{*}}}\right)^{\alpha+1} \exp \left(-\frac{L}{\mathrm{L^{*}}}\right) \ln 10 \mathrm{~d}(\log L)
\end{equation}
\begin{equation}
    \phi(M_{\star}) \mathrm{d} M=\mathrm{\phi^{*}}\left(\frac{M_{\star}}{\mathrm{M^{*}}}\right)^{\alpha+1} \exp \left(-\frac{M_{\star}}{\mathrm{M^{*}}}\right) \ln 10 \mathrm{~d}(\log M_{\star}),
\end{equation}
where $L$ is the \ha\ luminosity, $\mathrm{L^{*}}$ is a characteristic luminosity, $\mathrm{\phi^{*}}$ is the normalization constant, and $\alpha$ is the slope of the power-law on the low luminosity side.
$\mathrm{M^{*}}$ is a characteristic stellar mass, $\mathrm{\phi^{*}}$ is the normalization constant, and $\alpha$ is the slope of the power law at the low mass side.
We utilize the ecmee \citep{Foreman-Mackey2013}, which implements the affine-invariant ensemble sampler for MCMC proposed by \cite{Goodman2010}.
In the fitting process, we remove the faintest (lowest mass) bins where the number of galaxies is dramatically reduced around the limiting \ha\ luminosity and stellar mass.
Also, we remove the one brightest bin of USS1558 and the two brightest bins of PKS1138 in the \ha\ luminosity function.
Note that the confirmed AGNs have been removed from the \ha\ luminosity function (see \ref{lim_sample}), and we do not consider the Eddington bias correction \citep[e.g.,][]{Ilbert2013,Caputi2015,Grazian2015,Davidzon2017}.
The data points at the low-mass end have been corrected for completeness, but they are not as accurate as the brighter side, so we remove the two lowest mass bins in this fitting, following \cite{Shimakawa2018uss}.
The fitting results are shown in Table~\ref{table:3}.

\begin{figure*}
  \begin{tabular}{cc}
    \begin{minipage}{8.5cm}
      \begin{center}
        \includegraphics[width=\columnwidth]{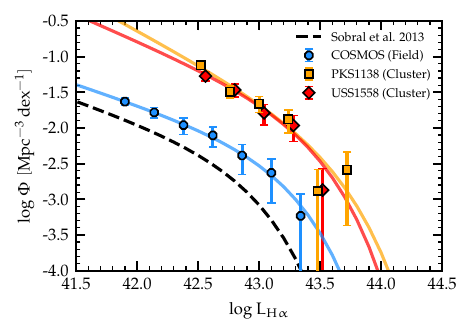}
      \end{center}
    \end{minipage}
    \begin{minipage}{8.5cm}
      \begin{center}
        \includegraphics[width=\columnwidth]{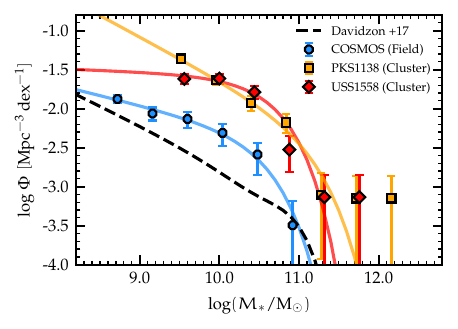}
      \end{center}
    \end{minipage}
  \end{tabular}
\vspace{-0.2cm}
\caption{Dust-corrected luminosity function (left) and stellar mass function (right).
Blue, red, and yellow curves correspond to COSMOS, USS1558, and PKS1138, respectively.
Dotted lines indicate the results for field galaxies in the literature.
Our COSMOS field functions show higher normalization compared to those in the literature, and our protocluster functions are even higher.}
\label{Fig:func}
\end{figure*}

\begin{figure*}
  \begin{tabular}{cc}
    \begin{minipage}{8.5cm}
      \begin{center}
        \includegraphics[width=\columnwidth]{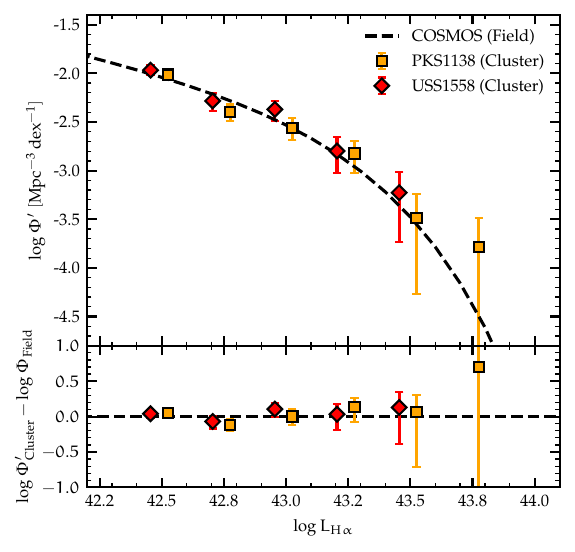}
      \end{center}
    \end{minipage}
    \begin{minipage}{8.5cm}
      \begin{center}
        \includegraphics[width=\columnwidth]{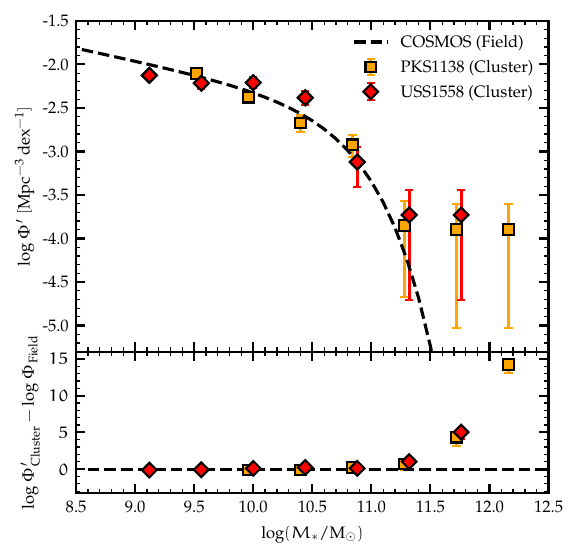}
      \end{center}
    \end{minipage}
  \end{tabular}
\vspace{-0.2cm}
\caption{Normalized Luminosity function (left) and stellar mass function (right). Blue, red, and yellow curves correspond to COSMOS, USS1558, and PKS1138, respectively. The $L_{\mathrm{H\alpha}}$ functions of USS1558 and PKS1138 are about 7 and 9 times higher than the $L_{\mathrm{H\alpha}}$ function of COSMOS, respectively, and $M_{\rm \star}$ functions of the protoclusters are about 5 times higher than the field function. The shapes of the $L_{\mathrm{H\alpha}}$ function are not different between the protoclusters and the COSMOS field. while the shapes of the $M_{\star}$ functions are different at the massive end.}
\label{Fig:norm_func}
\end{figure*}

\begin{table*}
 \label{table:3}
\centering
\caption{Fitting results for $L_{\mathrm{H\alpha}}$ and $M_{\rm \star}$ functions. We fix the slopes of the functions. The protocluster data are shallower than our COSMOS data, making it difficult for us to determine the slope. PKS1138 shows the steepest slope in spite of the fact that the data are the shallowest.}
\begin{tabular}{cccccccc}
\hline
Sample & Environment & $\log(L_{\mathrm{H\alpha}})$ erg s$^{-1}$ & $\log(\Phi^{*}_{\mathrm{H\alpha}})$ Mpc$^{-3}$ & $\alpha$ & $\log(M_{\star}^{*})$ M$_{\odot}$ & $\log(\Phi^{*}_{M_{\star}^{*}})$ Mpc$^{-3}$ & $\alpha$\\\hline
    COSMOS & Field & $43.15^{+0.21}_{-0.18}$ & $-2.67^{+0.25}_{-0.28}$ &  $-1.56^{+0.15}_{-0.14}$ & $10.61^{+0.20}_{-0.16}$ & $-2.72^{+0.18}_{-0.20}$ & $-1.25^{+0.09}_{-0.09}$\vspace{0.2cm}\\
     & & $43.14^{+0.16}_{0.13}$ & $-2.64^{+0.10}_{0.11}$ &  $-1.56$ (fix) & $10.61^{+0.13}_{0.11}$ & $-2.72^{+0.06}_{-0.06}$ & $-1.25$ (fix)\\\hline
    USS1558 & Cluster & $43.14^{+0.22}_{-0.22}$ & $-1.79^{+0.29}_{-0.36}$ &  $-1.45^{+0.45}_{-0.31}$ & $10.72^{+0.24}_{-0.17}$ & $-1.96^{+0.14}_{-0.16}$ & $-1.04^{+0.10}_{-0.09}$\vspace{0.2cm}\\
     &  & $43.22^{+0.12}_{0.09}$ & $-1.9^{+0.09}_{0.10}$ &  $-1.56$ (fix) & $10.87^{+0.37}_{0.20}$ & $-2.13^{+0.13}_{-0.16}$ & $-1.25$ (fix)\\\hline
    PKS1138 & Cluster & $43.26^{+0.26}_{-0.23}$ & $-2.08^{+0.41}_{0.51}$ &  $-1.81^{+0.35}_{-0.28}$ & $11.21^{+0.36}_{-0.25}$ & $-2.66^{+0.29}_{-0.37}$ & $-1.55^{+0.11}_{-0.10}$\vspace{0.2cm}\\
     &  & $43.09^{+0.16}_{0.12}$ & $-2.17^{+0.10}_{-0.13}$ &  $-1.56$ (fix) & $10.78^{+0.08}_{-0.06}$ & $-2.06^{+0.04}_{-0.04}$ & $-1.25$ (fix)\\\hline
\end{tabular}
\end{table*}

The $L_{\mathrm{H\alpha}}$ and $M_{\rm star}$ functions of the protocluster are more elevated than those of the field. 
Figure~\ref{Fig:norm_func} shows the normalized $L_{\mathrm{H\alpha}}$ and $M_{\rm star}$ function of the two protoclusters, which are scaled to match that of the field.
We have normalized them to minimize the difference between them.
As a result, we conclude that the $L_{\mathrm{H\alpha}}$ functions are about 7 and 9 times higher for the USS1558 and PKS1138, respectively, and $M_{\star}$ functions are about 5 times higher.
These factors ensure that the two protoclusters are much higher-density regions than the general field.
Our field luminosity function is also a bit higher than that in the literature \cite{Sobral2013}. 
This discrepancy may be due to the difference in limiting magnitude, dust extinction correction, \nii/\ha\ ratio and the cosmic variance.
Our N2 correction is typically 10\%, while their correction is around 25\% \citep{Sobral2013}. 
Thus, our luminosity function is systematically and slightly higher than \cite{Sobral2013} luminosity function.
The median limiting magnitude of the \cite{Sobral2013} is around 22.8 mag.
\cite{Sobral2013} assumed $A_V$ = 1 mag, while the $A_V$ median value of our target with $m_{\mathrm{NB}}$ $<$ 22.8 mag is 1.0 mag. Therefore, this assumption does not deviate significantly from our results. 
This means that, with regard to the bright side, differences in assumption of dust extinction do not have a significant impact.
When luminosity functions are constructed under the same conditions and compared, it is found that the bright ends match within one sigma errors.
This discrepancy may also be affected by the entry of previously invisible sources.
Although a part of the discrepancy can be attributed to the above possibilities, the offset may still remain, and we do not know yet if this is due to the intrinsic cosmic variance or any systematic difference between the two work.
In fact, our protocluster regions (USS1558 and PKS1138) show clear excesses ($>$ 8 times) in  $L_{\mathrm{H\alpha}}$ and $M_{\star}$ functions with respect to this field sample.
We regard our COSMOS emitter sample as the general field comparison sample.

The shape of the $L_{\mathrm{H\alpha}}$ functions for the protoclusters are both consistent with the field function, while the massive end of the $M_{\star}$ function show clear deviations for both protoclusters with respect to the field.
Such excess of massive emitters in protoclusters is consistent with the downsizing scenario \citep{Cowie1996,Kodama2004a} where galaxy formation and evolution take place earlier and proceed in an accelerated way in high-density regions.
These differences may also be due to environmental dependence in AGN activities.
On the other hand, we do not see a significant difference for the lower mass galaxies.
However, the fitting result shows that the slope of the stellar mass function of USS1558 is flatter than the other two samples.
This trend is also seen in \cite{Shimakawa2018uss}.
We note that our completeness correction may not be sufficient, but we stress that our completeness correction is stricter (see \ref{Completeness}) than that adopted in \cite{Shimakawa2018uss,Shimakawa2018pks}.
Deeper NB and Ks observations are needed to know if this steeper slope at the low-mass end is real or due to the observational limitations.

In contrast, PKS1138, despite of its shallowest data, shows the steepest slope $\alpha$ at the faint/low-mass end.
Recently, Sunyaev-Zel'dovich effect (SZE) has been detected in PKS1138 protocluster  \citep{Luca2023}.
SZE is caused by the inverse Compton scattering of the background CMB photons by the hot intracluster medium (ICM) associated with the protocluster core.
Therefore, it is possible that star-forming galaxies are produced by ram-pressure stripping in the protocluster core.
However, we are currently unable to discuss these galaxies quantitatively due to the insufficient depth of our data to study low-mass galaxies.

\subsection{star-forming main sequence at $z$ $\sim$ 2}
\label{Results:3}
\begin{figure*}
  \begin{tabular}{cc}
    \begin{minipage}{8.5cm}
      \begin{center}
        \includegraphics[width=\columnwidth]{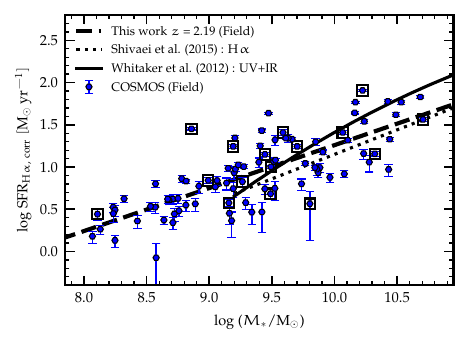}
      \end{center}
    \end{minipage}
    \begin{minipage}{8.5cm}
      \begin{center}
        \includegraphics[width=\columnwidth]{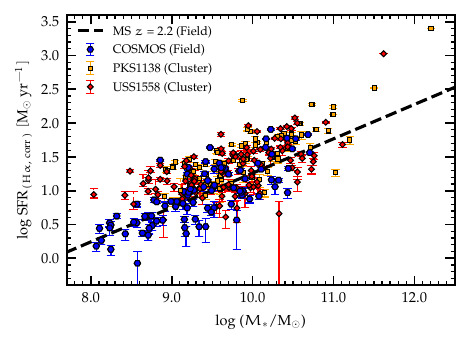}
      \end{center}
    \end{minipage}
  \end{tabular}
\vspace{-0.2cm}
\caption{Main sequence for the star-forming galaxies in the COSMOS field (left), and USS1558 and PKS1138 protoclusters (right). Blue data points indicate those in COSMOS, red ones indicate those in USS1558, and yellow ones indicate those in PKS1138, respectively. The thick dotted line in the left panel shows the best-fit line to our data. The solid and thin dotted lines show the results in previous studies \citep{Whitaker2012,Shivaei2015}. Black squares indicate spectroscopically confirmed emitters. Our results are consistent with the previous studies in the relatively massive regime. At the low-mass end ($M_{\star}<10^{9.0}\ \mathrm{M_{\odot}}$), however, many star-bursting galaxies exist in USS1558 which are located well above the main sequence of the field galaxies, which are missing in the general field. This suggests that some environmental effects may be at work for low-mass galaxies in high-density regions to enhance their star formation activities.}
\label{Fig:ms}
\end{figure*}

The star-forming main sequence (i.e., correlation between SFR and $M_{\star}$) is known to exist over wide ranges of stellar mass and redshift \citep[e.g.][]{Brinchmann2004,Daddi2007,Elbaz2007,Noeske2007,Salim2007,Magdis2010,Rodighiero2011,Wuyts2011,Whitaker2012,Whitaker2014,Heinis2014,Schreiber2015,Shivaei2015,Tomczak2016,Santini2017,Iyer2018,Leslie2020,Leja2022,Daddi2022}.
Figure~\ref{Fig:ms} shows the main sequence (MS) diagrams for our galaxy samples of field at $z$ = 2.19, PKS1138 at $z$ = 2.15, and USS1558 at $z$ = 2.53.
In this work, SFR is derived from \ha\ luminosity using NB and BB (see Equation~\ref{eq:lineflux}, \ref{eq_lineuss}, and \ref{eq_SFRha})
The Mahalo deep field survey has reached the depth corresponding to the stellar mass limit of $M_{\mathrm{lim,3\sigma}}\sim10^{8.5}\ \mathrm{M_{\odot}}$ and \ha-based dust-free star formation rate of SFR$_{\mathrm{lim,3\sigma}}$ $\sim$ 1.6 $\mathrm{M_{\odot}}$ yr$^{-1}$.
We estimate the limiting stellar mass by scaling the \cite{Kodama1998,Kodama1999} models assuming a passive evolution of stellar populations formed at $z_f$ = 5, following \cite{Hayashi2012,Hayashi2016}.
The SFR -- $M_{\star}$ correlations are established for all the samples over wide mass ranges. 
We perform a linear fit to the relation only using the emitters above $M_{\mathrm{lim,3\sigma}}\sim10^{8.5}
\mathrm{M_{\odot}}$, and obtained 
\begin{equation}
    \log_{10}({\rm SFR}) = 0.51 \log_{10}(M_{\star}/\mathrm{M_{\odot}}) - 3.82.
\end{equation}
At the mass range of $M_{\star}>10^{9.0}\ \mathrm{M_{\odot}}$, the SFR -- $M_{\star}$ relation is consistent with previous studies \citep[e.g.][]{Whitaker2015,Shivaei2015}.

On the other hand, the low mass end of the main sequence ($M_{\star}<10^{9.0}\ \mathrm{M_{\odot}}$) shows an intriguing discrepancy between the protocluster USS1558 and the general field (COSMOS).
In USS1558, we find 14 emitters at this low-mass end well above the main sequence, which can be called star-bursting galaxies. This trend is consistent with our previous result presented in \cite{Hayashi2016}.
Despite of the fact that our narrow-band imaging data for COSMOS field are deeper than those for USS1558 or PKS1138 protoclusters, we find only one such low-mass star-bursting galaxy in the general field region in COSMOS.

In this paper, we define $\Delta$MS as the deviation from the main sequence in the logarithmic of SFR;
\begin{equation}
\Delta\mathrm{MS}= \log_{10}(\mathrm{SFR}) - \log_{10}(\mathrm{SFR_{Field}}),
\end{equation}
which can be regarded as an indicator of enhancement in star formation activity, and we use the threshold of $\Delta \mathrm{MS}>0.3$ to identify star-burst galaxies since the typical dispersion of the main sequence is $\sim$ 0.3 dex \citep[e.g.][]{Nelson2016,Whitaker2012,Schreiber2015}.
The absolute value of the SFR may not be trustworthy since the dust extinction is derived from the SED fitting under the simple assumption of $E(B-V)_{\mathrm{gas}}=E(B-V)_{\mathrm{stellar}}$.
However, since we use the same photometric methods and SED fit parameters,
the relative comparison of the derived SFRs must be more robust.
It is intriguing that we find such many star-bursting galaxies in the young protocluster USS1558, even though the USS1558 data are shallower than the COSMOS data.
Star formation activity at the low-mass end will be further discussed in Section\ref{Discussion:1}.

\subsection{Environmental dependence at $z$ $\sim$ 2}

\subsubsection{Sample construction for comparison}
\label{lim_sample}
As the quality (depth) of the observed data differs significantly from field to field, a fairer comparison can be made only after adjusting the depth of the data.
For this purpose, we made a restricted sample where we use the uniform selection cuts for emission-line galaxies in the equations~(\ref{eq:SNcut}) and (\ref{eq:EWcut}) to adjust the depth of the emitter samples in COSMOS and USS1558 to match that of the shallowest sample in PKS1138.
Moreover, we select only the galaxies brighter than the 68\% completeness limit in NB and above the limiting stellar mass.

We stress that we apply the same methods and set the same criteria in photometries and sample selections for a fairer comparison.
It reduces any systematic errors due to different methods and selections, making our analyses more reliable.
The following subsections show the results from the comparisons of physical quantities based on these samples.

\subsubsection{Centrally concentrated star formation activity driven by environmental effects}
\label{Results:4}
We investigate star formation activities at the center of galaxies and compare them in three different environments.
Firstly, we perform aperture photometries within the PSF sizes and calculate star formation rates at galaxy centers using the same method.
We quantify the central concentration of star formation activities by taking the ratio of SFR at the center with respect to the total SFR represented by Kron magnitudes.
The larger this ratio is, the more compact, centralized star formation is taking place.
However, the SFR at the galactic center is underestimated because of higher dust extinction at galaxy centers than that in the outer regions. 
Note that compact objects such as low-mass galaxies are not spatially resolved.
We also investigate the galaxies using the mass-limited samples of $\log_{10}(M_{\star}/\mathrm{M_{\odot}})>9.2$, which corresponds to the shallowest limiting stellar mass of our samples for PKS1138. 

\begin{figure*}
	\includegraphics[width=17cm]{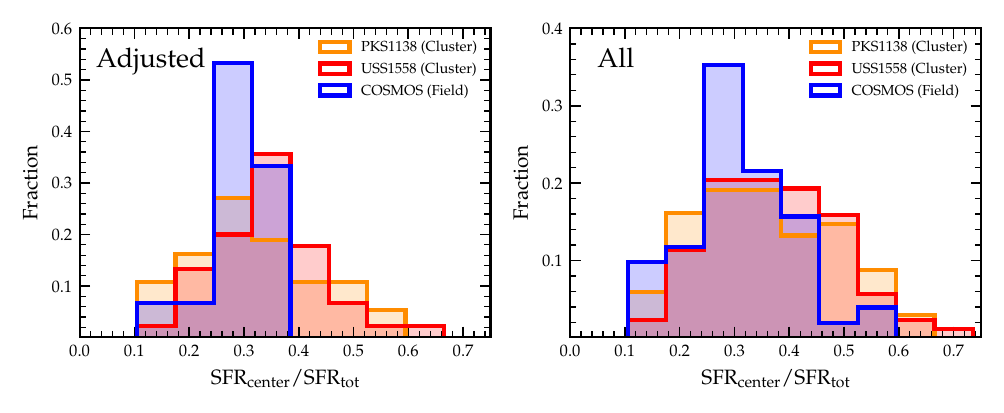}
    \vspace{-0.2cm}
    \caption{Compact star formation activities characteristic of protocluster galaxies. The left panel shows the limited sample with the adjusted selection criteria, while the right panel shows the entire sample. We quantify the compactness of star formation activity by taking the ratio of SFR within the PSF-sized (FWHM) aperture and the total SFR derived from the total magnitude. There is a significant difference in the distributions of the compactness of star formation between galaxies in protoclusters and those in the COSMOS (field), in the sense that protocluster galaxies tend to have more compact star formation activities. We note that this ratio of star formation rates would be underestimated because of the expected higher dust extinction towards galaxy centers. }
    \label{fig:hist}
\end{figure*}

Figure~\ref{fig:hist} shows that many galaxies with ratios greater than 0.4 are in the protoclusters.
The ratio distribution in PKS1138 and USS1558 is, in fact, significantly different from that in COSMOS, as indicated by the K-S test ($p$<0.05).
This means that many galaxies in the protoclusters have more compact, centralized star formation activity than those in the field.
It is likely to be related to the compaction events, which can be triggered by mergers, tidal interactions, and violent disk instabilities \citep[e.g.][]{Dekel2003, Hopkins2006, Zolotov2015, Dekel2014, Tacchella2016}.
It is naturally expected that those events can occur more frequently in dense environments with deeper gravitational potentials.

\begin{figure*}
	\includegraphics[width=16.2cm]{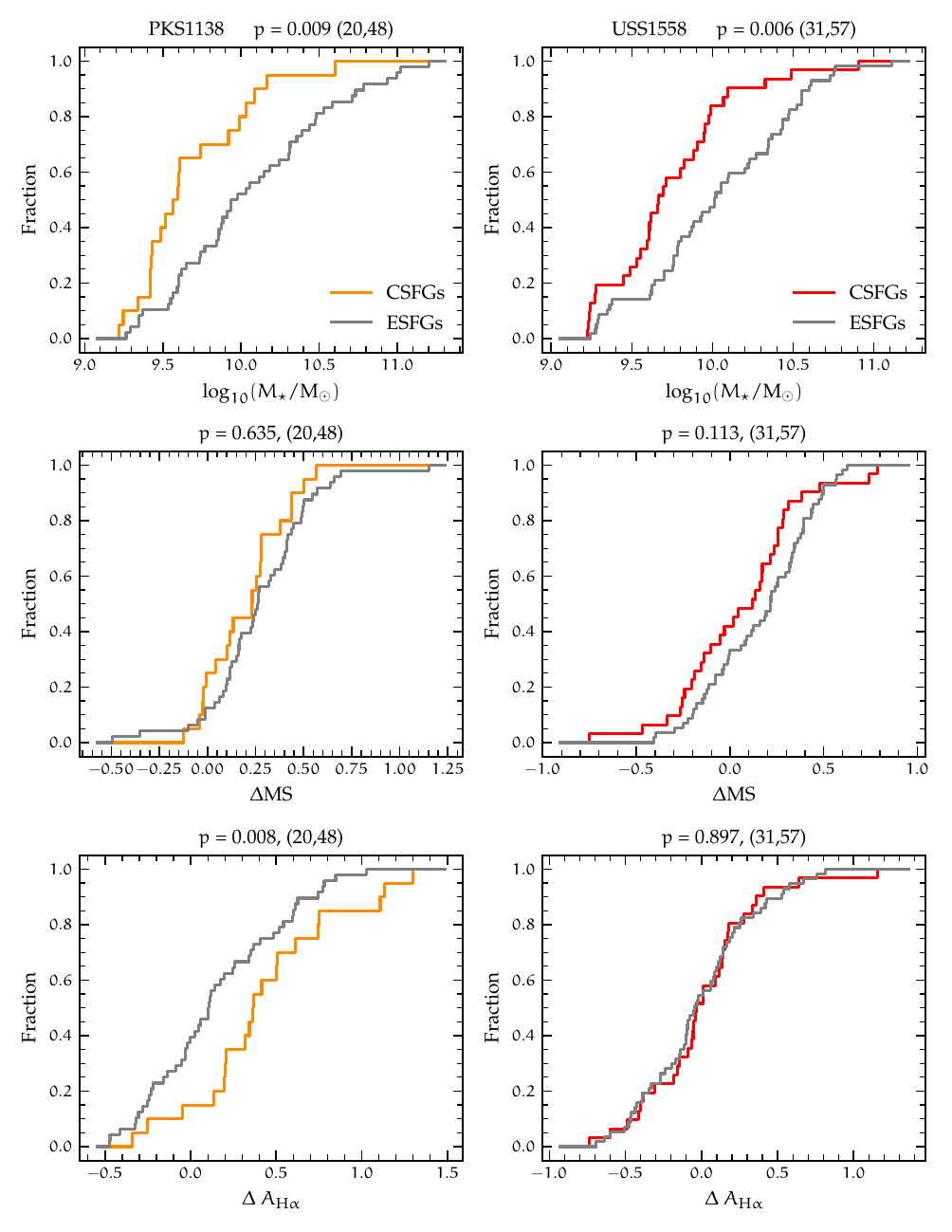}
    \vspace{-0.2cm}
    \caption{Cumulative distribution functions and the $p$-value derived from the K-S test to compare ESFGs and CSFGs. The top two panels show the stellar mass distribution of ESFGs and CSFGs. ESFGs are shown in grey color. CSFGs are shown in yellow (left) and red (right) color. The middle two panels show the $\Delta$MS distribution. The bottom two panels show the $\Delta A_{\mathrm{H\alpha}}$ ($= \mathrm{A_{H\alpha}} - \mathrm{A_{H\alpha,Field}}$) distribution.}
    \label{fig:comp_3}
\end{figure*}

We define the compact star-forming galaxies (CSFGs) and extended star-forming galaxies (ESFGs), as those having $\mathrm{SFR_{center}}/\mathrm{SFR_{total}}>0.4$ and $\mathrm{SFR_{center}}/\mathrm{SFR_{total}}<0.4$, respectively.
$\mathrm{SFR_{center}}$ is calculated from PSF size aperture flux using Equation~\ref{eq_line} and \ref{eq_SFRha}.
$\mathrm{SFR_{total}}$ consists with SFRs in Figure~\ref{Fig:ms}, which are derived from Kron magnitude.
CSFGs tend to be seen at the low-mass side compared to ESFGS, as shown by low $p$-values below 0.05 in the K-S test (Figure~\ref{fig:comp_3}). 
It suggests that compaction events are expected to occur more frequently in denser environments than in the field.

The \ha\ and continuum (stellar mass) distribution can be examined from the NB and BB images with Equation~\ref{eq_line} and \ref{eq_con}.
Due to their poor signal-to-noise ratio, we used stacked images here because not all galaxies can be measured accurately with these images.
The stacked images clearly indicate that CSFGs have centralized star formation activity.
We then directly measure the sizes of the galaxies from those stacked images using GALFIT \citep{Peng2002,Peng2010b}. 
The PSF images needed for the size measurements are also made using PSFEx \citep{Bertin2011}.
The \ha\ sizes (i.e. effective radii $R_{e,\mathrm{H}\alpha}$) of the CSFGs are 2.24 $\pm$ 0.02 kpc for USS1558 and 2.23 $\pm$ 0.09 kpc for PKS1138, respectively.
As for the ESFGs, the \ha\ sizes are 3.71 $\pm$ 0.03 kpc for USS1558 and 4.67 $\pm$ 0.07 kpc for PKS1138.
 Note that the PSF size corresponds to 2.9 kpc, and the \ha\ sizes of the CSFGs are a little bit smaller than the PSF size.
\cite{Aoyama2022} conducted CO(3-2) and dust continuum observations for USS1558 with ALMA at Band-3 and Band-6. 
One of the CSFGs, ID 38 \citep{Aoyama2022}, is detected with dust continuum emission.
The CSFG (ID38) has a compact size ($\mathrm{R_e}=1.47$ kpc in HST/WFC3 F160W), a low gas-to-stellar mass ratio, and a high star formation efficiency.
This object is likely to be in a transition phase to a passive galaxy, and a merger event may have caused a starburst at its center that has consumed a significant fraction of the gas.
Note, however, that not all the CSFGs are at this state, as they show various SFR distributions.
Nevertheless, this result is a piece of evidence supporting the scenario that the CSFGs are caused by galaxy margers/interactions.

Furthermore, the \ha\ sizes of the CSFGs tend to be smaller than the continuum sizes.
From the stacking analysis, the continuum size of the CSFGs in USS1558 is 3.28 $\pm$ 0.06 kpc.
Therefore the \ha\ sizes of the CSFGs are smaller than the continuum sizes ($R_{e,\mathrm{H}\alpha}<R_{e,\mathrm{con}}$).
This trend is consistent with the compaction scenario where the gas loses its angular momentum during galaxy interactions and falls towards galaxy centers.
Since the depth of the data is insufficient, we can not measure the continuum sizes of CSFGs in PKS1138.
On the other hand, ESFGs have more extended \ha\ distributions with respect to the continuum distributions, such as $R_{e,\mathrm{con}} = 4.67 \pm 0.07$ kpc for galaxies in USS1558 and $R_{e,\mathrm{con}} =  3.76 \pm 0.03$ kpc for those in PKS1138.
We note that \cite{Suzuki2019} compare the radial profiles of \ha\ emission line and the stellar continuum light based on a stacking analysis of star-forming galaxies in USS1558, and show that \ha\ tends to be more spatially extended than the continuum.
In field star-forming galaxies, \ha\ tends to be more extended than the stellar continuum \citep[][]{Nelson2016,Wilman2020}.
Our field samples also show a similar trend.
The continuum sizes of the CSFGs and ESFGs in USS1558 are similar (3.3kpc), as shown in Table~\ref{tab:size}.
The same continuum size for each object and the results shown in Figure~\ref{fig:hist} support the idea that the field galaxies are more like ESFGs.
Moreover, these results show that CSFGs in the protocluster, which shows a different trend, may be unique and/or characteristic to protoclusters.
\cite{Suzuki2019} and our result would differ partly because the majority of their sample are biased toward ESFGs (9/11).
Note that the method in \cite{Suzuki2019} is different, where they utilize high resolution images ($\sim$ 0.2 arcsec).

CSFGs are lower-mass systems compared to ESFGs, but their star-formation activities are consistent (Figure~\ref{fig:comp_3}).
The nature of the dust in CSFGs varies among protoclusters.
In the more-evolved protocluster PKS1138, CSFGs are found to be dustier systems than ESFGs.
On the other hand, in the younger protocluster USS1558, they are similar in dust extinction.
This difference may be due to differences in the evolutionary stage: PKS1138 is already at a more mature evolutionary stage than USS1558.
Thus, the dust extinction of PKS1138 may be larger than that of US1558.
These results suggest that the environment does matter in regulating star formation in galaxies and structure evolution of galaxies at $z\sim 2$.

\begin{table*}
\centering
\caption{Sizes (half-light radii) measured from the stacked images of compact star-forming galaxies (CSFGs) and extended star-forming galaxies (ESFGs). As defined, we can confirm that the \ha\ distribution of CSFGs tends to be more compact than that of ESFGs. Moreover, \ha\ distribution of CSFGs tends to be more compact than that of stellar distribution.}
\label{tab:size}
\begin{tabular}{cccc}\hline
    Sample &  & CSFG ($\mathrm{SFR_{center}}/\mathrm{SFR_{total}}>0.4$) [kpc] & ESFG ($\mathrm{SFR_{center}}/\mathrm{SFR_{total}}<0.4$) [kpc]\\\hline
    USS1558 & \ha       &  2.24 $\pm$ 0.02  & 3.71 $\pm$ 0.03 \\
    USS1558 & continuum &  3.28 $\pm$ 0.06  & 3.31 $\pm$ 0.03 \\
    PKS1138 & \ha       &  2.23 $\pm$ 0.09  & 4.67 $\pm$ 0.07 \\
    PKS1138 & continuum &  --                & 3.76 $\pm$ 0.03 \\ \hline
\end{tabular}
\end{table*}

\subsubsection{Enhanced star formation activities in the USS1558 protocluster}
\label{Results:5}
\begin{figure*}
	\includegraphics[width=16.2cm]{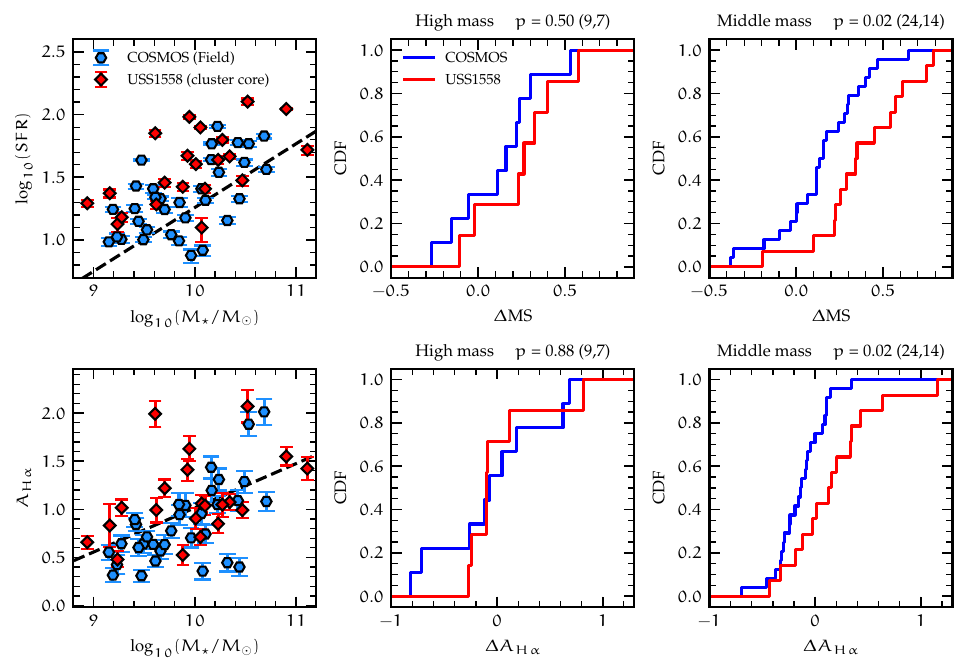}
	\includegraphics[width=16.2cm]{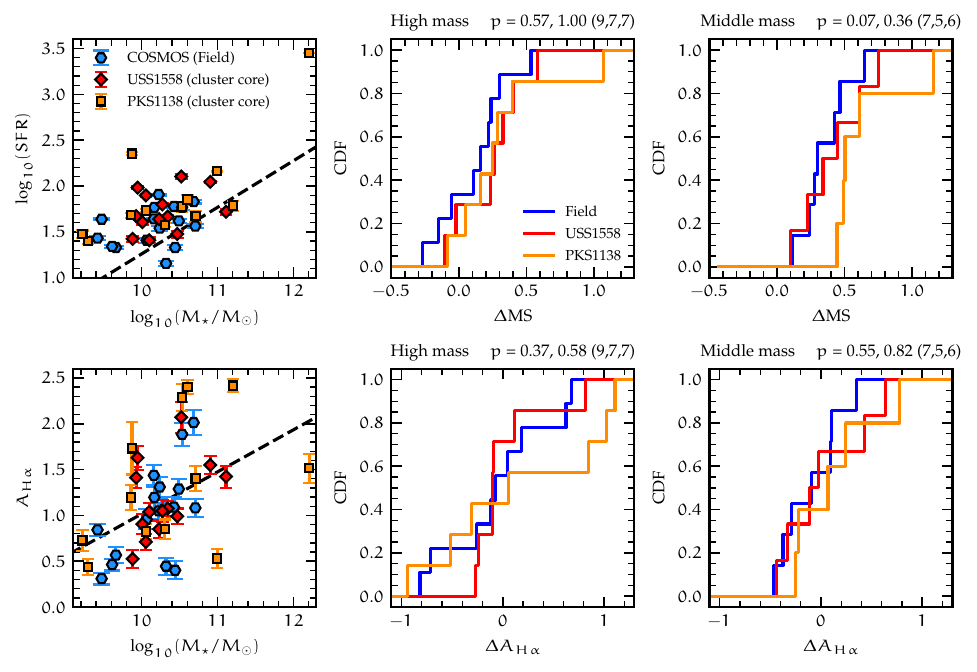}
    \vspace{-0.3cm}
    \caption{Cumulative distribution functions of some relative physical quantities and the $p$-values derived from the K-S test to compare galaxies in the field with those in the protocluster cores ($\bar{b}_{\mathrm{5th}}<150$ kpc). The top two panels show the restricted samples to match to the USS1558 data. The bottom two panels are the restricted sample to match to the PKS1138 data. The first and third rows compare $\Delta$MS, while the second and fourth rows show $\Delta A_{\mathrm{H}\alpha}$. The left most panels show the galaxy distributions on the SFR -- $\mathrm{M_{\odot}}$ and $A_{\mathrm{H}\alpha}$ -- $\mathrm{M_{\odot}}$ diagrams. The middle and the right panels show cumulative distribution functions for high and intermediate-mass galaxies separated at $10^{10.2}\ \mathrm{M_{\odot}}$. The $p$-values in the third and fourth rows show the results for COSMOS -- PKS1138 and USS1558 -- PKS1138 comparisons, respectively.}
    \label{fig:comp_1}
\end{figure*}

\begin{figure*}
	\includegraphics[width=16.2cm]{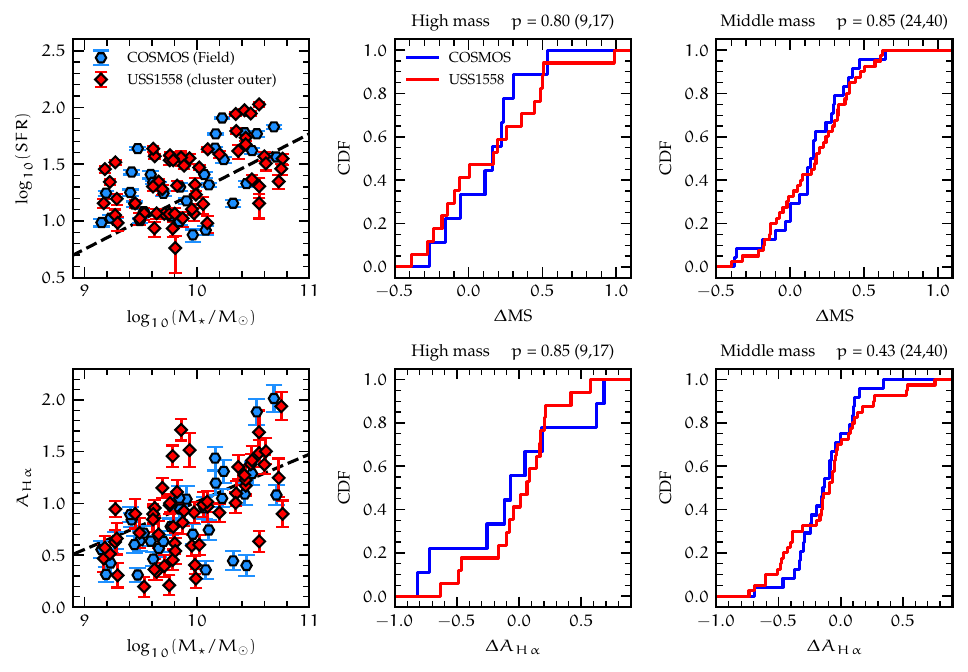}
	\includegraphics[width=16.2cm]{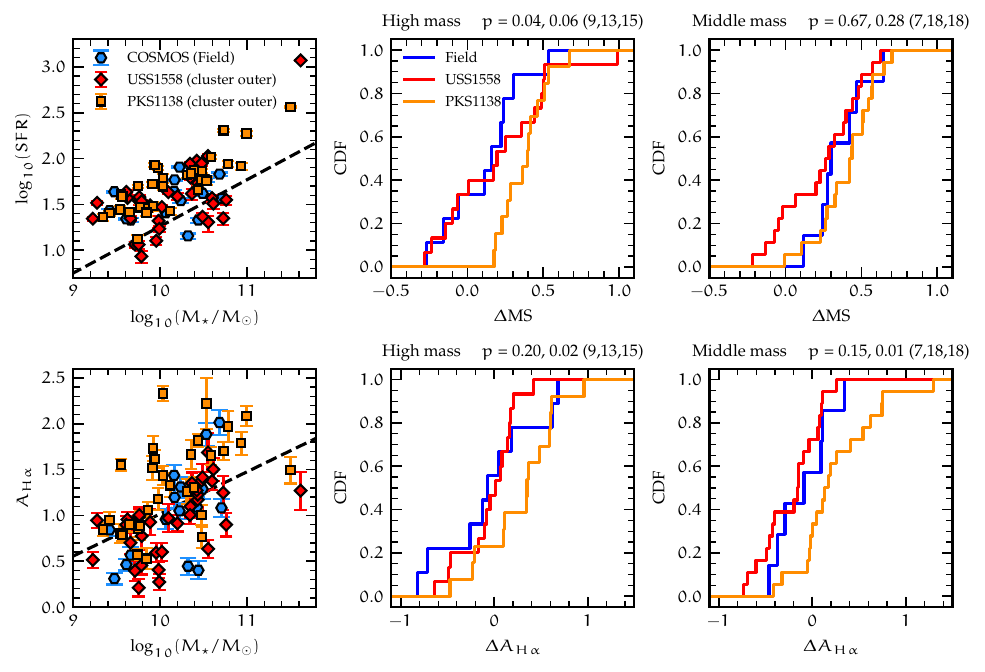}
    \vspace{-0.2cm}
    \caption{Cumulative distribution functions and the $p$-value derived from the K-S test to compare galaxies in the field and those in the protocluster outskirts ($\bar{b}_{\mathrm{5th}}>150$ kpc).}
    \label{fig:comp_2}
\end{figure*}

We investigate the environmental dependence of galaxy properties such as star formation rate, dust extinction, and stellar mass.
We quantify dust extinction with respect to the
A$_{\rm H\alpha}$ versus stellar mass relation for field galaxies, respectively, as defined by the following equations;
\begin{equation}
\Delta\mathrm{A_{H\alpha}}= \mathrm{A_{H\alpha}} - \mathrm{A_{H\alpha,Field}},
\end{equation}
where $\mathrm{SFR_{Field}}$ and $\mathrm{A_{H\alpha,Field}}$ are the corresponding values on the relationships for field galaxies at the same stellar mass.
We then compare (1) between the protocluster core regions ($\bar{b}_{\mathrm{5th}}<150$ kpc) and the field (Figure~\ref{fig:comp_1}), and (2) between the protocluster outer regions ($\bar{b}_{\mathrm{5th}}>150$ kpc) and the field (Figure~\ref{fig:comp_2}).
Figures~\ref{fig:comp_1} and \ref{fig:comp_2} show the SFR and $\mathrm{A_{H\alpha}}$ distributions of the restricted samples for the sake of fair comparisons (see \ref{lim_sample}).
In this study, star-forming galaxies with $M_{\star}>10^{10.2}\,\mathrm{M_{\odot}}$ are classified as massive galaxies, those with $10^{10.2}\gtrsim M_{\star}/\mathrm{M_{\odot}}>10^{8.9}\,\mathrm{M_{\odot}}$ are middle-mass galaxies, and those with $M_{\star}<10^{8.9}\,\mathrm{M_{\odot}}$ are low-mass star-forming galaxies.
For massive galaxies in the protocluster core, we do not identify any significant difference in $\Delta \mathrm{MS}$ distributions between the protoclusters and the field.
It is consistent with the previous studies in which no environmental dependence in star formation activities is found for massive galaxies \citep[][]{koyama2013,Jose2023}.
For intermediate-mass galaxies in the protocluster cores, however, USS1558 galaxies show statistically higher star formation activities ($\Delta \mathrm{MS}$) than the field galaxies ($p<0.05$), while no significant difference is seen in the case of PKS1138 galaxies (Figure~\ref{fig:comp_1}).
In the protocluster outskirts, there are no significant differences with respect to the field (Figure~\ref{fig:comp_2}) for both of the protoclusters.
In Figure~\ref{fig:comp_2}, the $p$-value for the $\Delta$MS distribution between COSMOS and PKS1138 is $p < 0.05$. 
However, if we slightly adjust the boundary between massive and intermediate-mass galaxies, the $p$-value of $\Delta$MS and $A_V$ distribution becomes $p > 0.05$ or $p < 0.05$. 
Hence, we cannot conclude that there is a significant difference in the star formation activity between PKS1138 and COSMOS based on the current data.
We note that we vary the boundary between massive and intermediate-mass galaxies over the range of $10^{10.0}$ -- $10^{10.3}$, and confirm that the other results do not change significantly. 
We conclude, therefore, that the star formation activity is enhanced only in the intermediate-mass galaxies in the protocluster core of USS1558.
The finding that the USS1558 core shows an enhanced star formation activity is consistent with the earlier work by \citep[][]{Shimakawa2018uss}.
As we discussed in \S4.4.2, the existence of CSFGs in the intermediate-mass
suggests that galaxy interactions/mergers and/or violent disk instability may be at work, and this is likely related to the enhancement of star formation.

\subsubsection{Inside-out star formation activities at $z\sim 2$}

Although there is no environmental dependence in star formation rates in the protocluster outskirts, there is a significant difference in the K-S test in dust extinction values. 
In fact, PKS1138 outskirts hosts more dusty galaxies than the USS1558 outskirts (Figure \ref{fig:comp_2}). 
PKS1138 shows a large, single structure centered on the radio galaxy, while USS1558 is composed of multiple clumps.
PKS1138 is expected to be a more evolved system than USS1558 in terms of the stellar mass function and the spatial distribution of member galaxies.
Note that we also need to investigate quiescent galaxies to confirm that PKS1138 is actually a more evolved system.
The results that there are more dusty galaxies in outside of PKS1138 than USS1558 suggest an inside-out evolution of clusters.
In PKS1138, the over-density region of the sub-millimeter galaxies does not coincide with the the central radio galaxy \citep{Dannerbauer2014}.
In USS1558, however, the sub-millimeter galaxies do exist at the center of the \ha\ emitter over-densities \citep{Aoyama2022}.
These results may support the inside-out growth scenario.
This trend has already been suggested at a lower redshift of $z\sim1$ \citep{koyama2013}.

PKS1138 data are the shallowest among our samples, which does not allow us to investigate the properties of galaxies with lower stellar masses or lower SFRs.
One of our future directions will be to investigate the properties of lower mass galaxies in PKS1138, which can be more sensitive to the environmental effects, as is seen at lower redshifts.

\section{Discussion}
\subsection{Enhanced star formation activity at the low-mass end in the protocluster}
\label{Discussion:1}
We find many galaxies in the protocluster USS1558 just below the limiting stellar mass ($\sim10^{9.0}$ $\mathrm{M_{\odot}}$), which are located well above the field MS as first reported by \cite{Hayashi2016}.
In contrast, we find only a few such galaxies in the field.
However, we cannot tell whether there are also many galaxies in USS1558 which are located on the field MS because the depth of the USS1558 data is shallower compared to the field data to make a fair comparison of the entire galaxies down to the corresponding SFR and stellar mass on the MS.
In fact, low-mass galaxies in USS1558 are not fully sampled because the mass ranges of low-mass galaxies in the USS1558 and PKS1138 are below the limiting stellar mass.

Therefore, we take the completeness into account as follows;
\begin{equation}
  N_{\mathrm{corr}} = \sum_i\frac{1}{C_i(m_{\mathrm{NB}})}.
\end{equation}
We first count the numbers of star-burst and whole galaxies at $\sim10^{8.9}$  $\mathrm{M_{\odot}}$ in the general field COSMOS, where the data are deep enough and complete.
The fraction of star-burst galaxies at this low-mass range is estimated to be $f_{\mathrm{burst}} = 0.02$.
Assuming that $\Delta$MS distribution has no environmental dependence, we can estimate how many low-mass galaxies are expected to lie on the MS.
In other words, we can estimate how many galaxies are expected to lie above the MS for the given total number of galaxies.
The total number of low-mass galaxies in this range estimated from the total stellar mass function of USS1558 is 413.4.
Therefore, the low-mass star-burst galaxies can be estimated to be 4.5, assuming $f_{\mathrm{burst}} = 0.02$.
Here we use the stellar mass function fitted with the fixed faint-end slope of $\alpha$.
Therefore, the actual observed number of low-mass starburst galaxies is 3.6 times larger than expected.
If the alpha is set free, the difference becomes even larger, multiplied by a factor of 7.1.
This result indicates that some environmental effects are at work at the lower mass side than $10^{8.9}\,\mathrm{M_{\odot}}$. 

The environmental influences on low-mass galaxies have been investigated in the literature.
\cite{Jeon2022} investigate quenching fraction and e-folding timescales ($\tau$) of star formation using the large-scale cosmological hydrodynamic simulation, Horizon-AGN.
They compare $\tau$ values between field and clustered galaxies and claim that the environmental effects are more significant for the galaxies that have existed longer in clusters.
Recently infalling galaxies have small $\tau$ value.
The environment may have already affected these galaxies as they pass through the dense environments before falling into the cluster \citep{DeLucia2012}.
The quenching mechanisms which are at work for low-mass galaxies can include stellar feedback (SN explosions) and ram pressure stripping \cite{Bekki2003}.
However, it is not clear if the ram pressure stripping can be efficient in high-$z$ protoclusters where the density may not be high enough.
Ram pressure can also enhance star formation \citep{Bekki2003} due to compression of the gas, but it is unclear if this is the origin of the enhanced star formation activity in the USS1558.
In addition to these quenching mechanisms, there are also some mechanisms which can elevate star formation activities. 
Those galaxies under the influence of activation/quenching can go up and down around the star-forming main sequence as discussed in \cite[e.g.][]{Tacchella2016}.
In any case, there have been quite limited pieces of information on low-mass galaxies in high-redshift protoclusters, and more observations and investigations are needed.

\subsection{Physical processes behind the environmental effects}
\label{Discussion:2}
We have demonstrated that some galaxies in the protocluster have compact star formation at the galaxy centers, as indicated by smaller \ha\ emission sizes compared to the stellar component sizes.
As we discussed in section~\ref{Results:4}, these trends may be caused by (i) galaxy-galaxy mergers/interactions and/or (ii) violent disk gas instability fed by cold stream accretions \citep[e.g.][]{Dekel2003, Hopkins2006, Zolotov2015, Dekel2014, Tacchella2016}.
Both of these events would drive the gas which lose angular momentum towards galaxy centers, and enhance the centralized star formation activities \cite{Kaviraj2013,Pearson2019,Shah2022}.
\cite{Naufal2023} investigate the rest-frame UV size and morphology in protoclusters at $z\sim2$ including PKS1138 and USS1558, suggesting that giant clumps and galaxy mergers caused the high star formation activity.
This observation also supports this scenario.

protoclusters at high-$z$ are indeed preferred sites of galaxy mergers/interactions due to their high number density of galaxies and relatively low-velocity dispersion.
\cite{Silva2018} shows that the fraction of 
star formation in low-mass galaxies is enhanced for a short period of time during major mergers \citep{Davies2015} and more dramatically affected than its companion with the masses of the nearby major mergers at distances of 3 -- 15 kpc increase \citep{Silva2018}.
However, it is yet unclear whether the mergers/interactions at high redshifts can really enhance star formation.
Some theories \citep[e.g.][]{Fensch2017,Martin2017,RodriguezMontero2019,Patton2020} and observations \citep{Shah2022} have noted that mergers/interactions at high redshifts may not enhance star formation activities to the same extent as in nearby galaxies.
However, in the high-$z$ universe ($z$ $>$ 5), low-mass galaxies tend to experience galaxy interactions more frequently \citep{Asada2023}, which may also trigger intense star formation activities.
Also, at $z>1$ protoclusters, enhanced merger fraction have been reported by many studies \citep{Lotz2013,Coogan2018,Watson2019,Hine2016,Liu2023}.
In particular, \cite{Lotz2013,Coogan2018,Watson2019,Liu2023} find 2–10 times higher merger fractions compered to the field at $z\sim2$.
On the other hand, there are some studies that argue the opposite trend \cite{Delahaye2017,Monson2021}.
Such discrepancy may be due to differences in dynamical state, method and/or sample size.
Moreover, we should also consider that not just drastic mergers but mere galaxy-galaxy interactions may also trigger the enhanced star formation activities in low-mass galaxies at high redshifts.
Therefore, we would argue that the frequency of mergers and/or interactions may be higher in the young protocluster USS1558 at $z\sim2.5$, which may lead to the environmental dependence that we see.

Difference in gas accretion process may be another cause of environmental variation of galaxy properties.
Gas accretion onto a galaxy is essential to continue star formation in the galaxy.
There are two modes of gas accretion, namely hot and cold modes.
In the hot mode, gas infall is accelerated by gravity in the deep potential well, and heated by shocks, and then cooled down slowly by radiation.
On the other hand, in the cold mode, fresh cold gas flows along the surrounding filamentary structures towards the galaxy at the intersection and accrete directly onto the galactic disk.
\cite{Keres2005,Keres2009} shows the fraction of cold mode gas accretion to the total gas accretion as a function of galaxy mass and redshift.
In protocluster galaxies, gas is expected to accrete to galaxies more efficiently in the cold mode. The gas-rich disk would become gravitationally unstable and fragmented into massive clumps, which would then be driven to the galaxy center due to dynamical friction, leading to the central starbursts.
We note, however, that it is difficult to resolve internal star formation within low-mass galaxies due to limited spatial resolution from the ground.

For massive galaxies with $M_{\star}>10^{10}\ \mathrm{M_{\odot}}$, we find no environmental dependence on star formation activities, in agreement with previous studies \citep[e.g.][]{koyama2013}.
This indicates that massive galaxies tend to quench star formation with a mechanism independent of the surrounding environment.
One possibility is the AGN feedback, although AGN fraction could also depend on the environment.
\cite{macuga2019} investigate the AGNs in the general field region and report that the fraction of AGNs is $4.8^{+2.0}_{-1.4}\%$ ($N=10/210$).
This suggests that the USS1558 protocluster has a lower fraction of AGNs than the field.
However, it is reported in the literature that protoclusters at $z>2$ tend to have higher fractions of AGNs than the contemporary field \cite{Digby-North2010,Lehmer2013,Krishnan2017,Monson2021,Tozzi2022}.
In any case, at this stage it is impossible to make any statistical argument based on the current sample constructed from a small area since AGNs are rare objects.
The environmental dependence of AGNs still needs to be explored and better understood in the future.
Identifying more protoclusters with deep X-ray and FIR observations is essential.

The absence of enhancement in star formation activities in PKS1138 may be because PKS1138 is a more evolutionary advanced system than USS1558, and thus can provide more environmental quenching processes to suppress star formation activities.
In USS1558, if we restrict the galaxy sample to the depth of PKS1138, there is no significant environmental dependence either.
This may be due to the small sample size of field galaxies, or that environmental dependence is seen more in lower-mass galaxies.
The quenching mechanism due to the environment also works for more massive galaxies.
Therefore, no difference in star formation activities at the massive end is identified between the protoclusters and the field.
We should also note that there is a significant scatter in the properties of protoclusters and galaxies therein even if they are located at similar redshifts.
We need to construct a large statistical sample of protoclusters of various ages, masses, and richness.

\section{Conclusions}
We have performed deep narrow-band imaging observations for the COSMOS field using MOIRCS on Subaru Telescope.
Our survey has reached the limiting magnitude down to 25.5 mag (1.5 arcsec diameter aperture, 3$\sigma$). 
We have identified 99 \ha\ emitters ($z=2.19$), 56 \oiii\ emitters ($z=3.16$), and 3 \oii\ emitters ($z=4.6$). 
In this study, we focus on the \ha\ emitters at $z\sim2$.
We examine the properties of $z\sim2$ star-forming galaxies over a wide mass range by separating the galaxies as a function of mass and environment.
We also compare this field sample to two protoclusters (USS1558 and PKS1138) we studied in our previous works to investigate environmental dependence in star-forming galaxies at $z\sim2$.

One of the key points of this study is that we reanalyze all the protocluster data using the same methods as we use for the field galaxies, including photometry, sample selection, and SED fitting, to make a fairer comparison\footnote{Note that we therefore compile a different emitter catalog of the protoclusters from those in the previous works.}.
This reanalysis reduces systematic errors due to differences in analysis methods.
In addition, since we are studying \ha\ line emitters as good tracers of star-forming galaxies, our analysis is more unbiased compared to, for example, \lya\ emitters, which are more prone to dust extinction and resonant scattering.
Our conclusions are summarized below.

We confirm the existence of many low-mass galaxies ($M_{\star}<10^{8.9}\,\mathrm{M_{\odot}}$) above the MS in USS1558 ($z=2.53$), as reported by \cite{Hayashi2016}.
At that time, \cite{Hayashi2016} could not tell whether these galaxies universally exist in all environments or they are seen only in protoclusters (or even only in USS1558) due to the lack of comparable field samples.
This study infers that these objects are unique to protoclusters since the number of star-bursting low-mass galaxies is 4-7 times larger than the expected number derived from the mass function and the bursting fraction of the field sample, if there is no environmental dependence.
We should admit, though that our data on the protoclusters are still insufficient to study the lowest-mass galaxies, and deeper observations are needed in the future.

For massive galaxies ($M_{\star}>10^{10.2}\,\mathrm{M_{\odot}}$), no environmental dependence is found in star formation rate and dust extinction distributions between the general field and the protoclusters.
However, intermediate-mass galaxies in the core of protocluster USS1558 do show environmental dependence in both of these quantities.
These results indicate that some environmental effects do at work in the protocluster core to promote star formation activities.
The reason why no difference is seen in the massive galaxies is probably because there are also other mechanisms in the protocluster core to suppress star formation activities.
In the protocluster PKS1138, no environmental dependence is seen for both high and intermediate-mass galaxies.
This may be because PKS1138 is a more evolved system than USS1558, and the star formation suppression mechanism is more effective even for intermediate-mass galaxies.
However, we should keep in mind that no significant difference is actually seen even for USS1558 if the sample is limited to match the depth of PKS1138 data.
Therefore, it is yet unclear if the environmental effects do exist or not until more statistical samples are constructed with a larger number of protoclusters.
We will then be able to investigate galaxy properties over a broader range in mass and SFRs with greater statistics to obtain more significant results.  

In the outskirts of PKS1138, there are more dusty galaxies than in USS1558.
This may suggest the inside-out star formation in clusters, where even the galaxies in the outskirts of PKS1138 are in an advanced stage compared to USS1558.
We also note that our current field galaxy sample do not have enough number of massive galaxies.
Therefore, a much wider field survey must be performed in order for us to make more statistically significant comparison with the protoclusters.

The viable mechanisms of the enhanced star-formation activities observed in low and intermediate-mass galaxies in the protocluster cores are likely to be a higher frequency of galaxy mergers/interactions and/or higher gas accretion rates.
We also identify some galaxies in the protoclusters with compact star formation activities, and their sizes of star-forming regions are more compact than the sizes of underlying older stellar distributions.
This result may be related to the compaction event, where the gas can lose its angular momentum due to mergers/interactions or violent disk instabilities, and it flows toward galaxy centers leading to compact starbursts. 
This effectively makes the half-light sizes of star-forming regions smaller.
The protocluster cores are located at the crossroads of the cosmic filamentary structures and thus can accrete the surrounding gas more efficiently.
Moreover, the protoclusters are ideal sites for mergers/interactions.
We propose that the combination of these factors may promote star formation activities in young protocluster environments.

We have shown that the environmental effects seem to be at work to enhance star formation activities at the low-mass regime.
Deeper observations are needed to further discuss these intriguing results with greater accuracy.
The low-mass star-forming galaxies in high-redshift primordial clusters have yet to be well studied.
Our current samples, both in the field and protoclusters, are likely to be affected by cosmic variance.
The recently launched satellite Euclid will be powerful in finding new protoclusters and will thus provide us with more statistical samples of proto-lcusters for environmental studies in the cosmic noon. Furthermore, JWST or next generation ground-based projects such as ULTIMATE-Subaru and ELTs (including TMT), will probe the galaxy properties to lower masses to confirm the results presented in this work, and to extend further down the low-mass end of the stellar mass functions.

\section*{Acknowledgements}
This research is based on data collected at the Subaru Telescope, which is operated by the National Astronomical Observatory of Japan. We are honored and grateful for the opportunity of observing the Universe from Maunakea, which has the cultural, historical, and natural significance in Hawaii.
This work is partly supported by JST SPRING, Japan Grant Number JPMJSP2114, and Graduate Program on Physics for the Universe (GP-PU), Tohoku University.
TK acknowledges JSPS KAKENHI Grant Number 18H03717 for financial support.
This work is also supported by JSPS KAKENHI Grant Number 22K21349 (International Leading Research).
JMPM acknowledges that this project has received funding from the European Union’s Horizon research and innovation programme under the Marie Skłodowska-Curie grant agreement No 101106626.
The Hyper Suprime-Cam (HSC) collaboration includes the astronomical communities of Japan and Taiwan, and Princeton University. The HSC instrumentation and software were developed by the National Astronomical Observatory of Japan (NAOJ), the Kavli Institute for the Physics and Mathematics of the Universe (Kavli IPMU), the University of Tokyo, the High Energy Accelerator Research Organization (KEK), the Academia Sinica Institute for Astronomy and Astrophysics in Taiwan (ASIAA), and Princeton University. Funding was contributed by the FIRST program from Japanese Cabinet Office, the Ministry of Education, Culture, Sports, Science and Technology (MEXT), the Japan Society for the Promotion of Science (JSPS), Japan Science and Technology Agency (JST), the Toray Science Foundation, NAOJ, Kavli IPMU, KEK, ASIAA, and Princeton University. 
This paper makes use of software developed for the Large Synoptic Survey Telescope. We thank the LSST Project for making their code available as free software at  http://dm.lsst.org
The Pan-STARRS1 Surveys (PS1) have been made possible through contributions of the Institute for Astronomy, the University of Hawaii, the Pan-STARRS Project Office, the Max-Planck Society and its participating institutes, the Max Planck Institute for Astronomy, Heidelberg and the Max Planck Institute for Extraterrestrial Physics, Garching, The Johns Hopkins University, Durham University, the University of Edinburgh, Queen’s University Belfast, the Harvard-Smithsonian Center for Astrophysics, the Las Cumbres Observatory Global Telescope Network Incorporated, the National Central University of Taiwan, the Space Telescope Science Institute, the National Aeronautics and Space Administration under Grant No. NNX08AR22G issued through the Planetary Science Division of the NASA Science Mission Directorate, the National Science Foundation under Grant No. AST-1238877, the University of Maryland, and Eotvos Lorand University (ELTE) and the Los Alamos National Laboratory.
Based in part on data collected at the Subaru Telescope and retrieved from the HSC data archive system, which is operated by Subaru Telescope and Astronomy Data Center at National Astronomical Observatory of Japan.
This work is based in part on observations taken by the 3D-HST Treasury Program (GO 12177 and 12328) and the CANDELS Multi-Cycle Treasury Program with the NASA/ESA HST, which is operated by the Association of Universities for Research in Astronomy, Inc., under NASA contract NAS5-26555.

Software: Astropy \citep{astropy:2013, astropy:2018, astropy:2022}, Numpy \citep{harris2020array}, Scipy \citep{2020SciPy-NMeth}, Photutils \citep{Photutils}, EMCEE \citep{Foreman-Mackey2013}.

\section*{Data Availability}
The Subaru photometric data is utilized through the Subaru Mitaka Okayama Kiso Archive (SMOKA) system.
Our NB-selected sample of star-forming galaxies can be shared on reasonable request.



\bibliographystyle{mnras}
\bibliography{main.bib} 




\appendix


\bsp	
\label{lastpage}
\end{document}